\documentclass[useAMS,usenatbib,fleqn]{mnras}
\usepackage[utf8]{inputenc}
\usepackage{amsmath} \usepackage{caption} \usepackage{amsfonts}
\usepackage{amssymb} \usepackage{graphicx} \usepackage{hyperref}
\usepackage{subfigure} \usepackage[crop=pdfcrop]{pstool} \usepackage{bm} \usepackage{multirow}
\usepackage{rotating} 
\usepackage[dvipsnames]{xcolor}
\usepackage[most]{tcolorbox}
\usepackage{scalerel}
\usepackage[normalem]{ulem}

\def\vis{\mathcal{V}\left( \mathbfit{U}, \nu \right)}

\def\visnonu{\mathcal{V}\left( \mathbfit{U} \right)}
\def\signonu{\mathcal{S}\left( \mathbfit{U} \right)}
\def\nsnonu{\mathcal{N}\left( \mathbfit{U} \right)}
\def\visa{\mathcal{V}\left(\alp, \mathbfit{U}, \nu \right)}

\def\v2{V_2\left(\mathbfit{U},\Delta \nu = 0\right)}

\def\dBdTnu{Q_{\nu}}

\newcommand{\vcg}{\mathcal{V}_{cg}}

\def\HIns{{\ion{H}{I}}}
\def\HI{{\ion{H}{I}~}}

\def\n{\hat{\mathbfit{n}}} 
\def\m{\hat{\mathbfit{p}}}
\def\dn{\Delta \mathbfit{n}}

\newcommand{\uv}[1]{\hat{\mathbfit{#1}}}

\newcommand{\bk}{\mathbfit{k}}

\newcommand{\alp}{\alpha_p}
\newcommand{\alc}{\alpha_c}
\newcommand{\tht}{\theta}
\newcommand{\thw}{\theta_w}
\newcommand{\thf}{\theta_{\rm FWHM}}
\newcommand{\thta}{\theta_1}
\newcommand{\thtb}{\theta_2}
\newcommand{\vtht}{\bm{\theta}}
\newcommand{\vch}{\bm{\chi}_p}
\newcommand{\vthtp}{\bm{\theta}-\bm{\chi}_p}
\newcommand{\U}{\mathbfit{U}}
\def\cl{C_{\ell}}

\def\V{\mathcal{V}}

\def\eg{{\it e.g.}\,}
\def\ie{{\it i.e.}\,}

\def\u{{\bf U}} 

\def\lsim{~\rlap{$<$}{\lower 1.0ex\hbox{$\sim$}}}

\def\gsim{~\rlap{$>$}{\lower 1.0ex\hbox{$\sim$}}}

\newcommand\redst{\bgroup\markoverwith{\textcolor{red}{\rule[0.5ex]{2pt}{0.4pt}}}\ULon}

\title[Tracking Tapered Gridded  Estimator ]{The Tracking Tapered Gridded Estimator (TTGE)  for the power spectrum from drift scan observations}

\author[Chatterjee, et al.]{
 Suman Chatterjee$^{1}$, Somnath Bharadwaj$^{2}$,  Samir Choudhuri$^{3,4}$,   Shiv Sethi$^{5}$ and Akash Kumar Patwa$^{5}$\\
$^{1}$ National Centre for Radio Astrophysics, Tata Institute of Fundamental Research, Post Bag 3, Ganeshkhind, Pune - 411 007, India.\\
$^{2}$ Department of Physics, Indian Institute of Technology Kharagpur, Kharagpur - 721 302, India.\\
$^{3}$ Astronomy Unit, Queen Mary University of London, Mile End Road, London E1 4NS, United Kingdom.\\
$^{4}$Centre for Strings, Gravitation and Cosmology, Department of Physics, Indian Institute of Technology Madras, Chennai 600036, India \\
$^{5}$ Raman Research Institute, C. V. Raman Avenue, Sadashivanagar, Bengaluru 560080, India.
}

\date{Accepted XXX. Received YYY; in original form ZZZ}
\pubyear{2022}

\begin{document}
\label{firstpage}
\pagerange{\pageref{firstpage}--\pageref{lastpage}}
\maketitle

\begin{abstract}
Intensity mapping with the redshifted 21-cm line is an emerging tool in cosmology. 
Drift scan observations,  where  the antennas are fixed to the ground and  the telescope's pointing center (PC) changes continuously on the sky due to earth's rotation, provide  broad sky coverage and sustained instrumental stability needed for 21-cm intensity mapping. Here  we present the Tracking Tapered Grided Estimator (TTGE) to quantify the power spectrum of the sky signal estimated directly from the visibilities measured in  drift scan radio interferometric observations.  The TTGE uses the data from the different PC to estimate the power spectrum of  the signal from a small angular region located around  a  fixed tracking  center (TC).  The size of this angular region is decided by a suitably chosen tapering window function which serves to reduce the foreground contamination from bright sources located at large angles from the TC. It is possible to cover the angular footprint of the drift scan observations using multiple TC, and combine the estimated power spectra to increase the signal to noise ratio.  Here  we have validated the  TTGE using simulations of $154 \, {\rm MHz}$ MWA drift scan observations. 
 We show that the TTGE can recover the input model angular power spectrum $\cl$  within $20 \%$ accuracy over the $\ell$ range $40 < \ell < 700$. 
 
\end{abstract}



 \section{Introduction}\label{sec:intro}
Radio-interferometric observations of the redshifted $21$-cm radiation  from neutral Hydrogen (\HIns) hold the potential to probe the Universe across a large redshift ($z$) range  from the Dark Ages $z \sim 50$ all the way to the Post-reionization Era $z <6$ \citep{Bharadwaj2005}. The 21-cm signal from the Epoch of Reionization (EoR), when the diffuse \HI in the intergalactic medium was ionized by the first galaxies and quasars, is of particular interest  \citep{Furlanetto2006,Morales2010,Pritchard2012,Mellema2013}. Several radio interferometers such as
 the Giant Meterwave Radio Telescope (GMRT\footnote{\url{http://www.gmrt.ncra.tifr.res.in/}}; 
 \citealt{Swarup1991}), the Low Frequency Array (LOFAR\footnote{\url{http://www.lofar.org/}}, \citealt{vanHarlem2013}),  
the Murchison Wide-field Array (MWA\footnote{\url{http://www.mwatelescope.org}} \citealt{Tingay2013}), 
the Donald C. Backer Precision Array to Probe the Epoch of Reionization (PAPER\footnote{\url{http://eor.berkeley.edu/}}, \citealt{Parsons2010}) and the Hydrogen Epoch of Reionization Array (HERA\footnote{\url{http://reionization.org/}}; \citealt{DeBoer2017}) are all involved in efforts  to observe the EoR 21-cm signal. Several  more sensitive upcoming telescopes such as the Square Kilometer Array (SKA\footnote{\url{http://www.skatelescope.org/}}; \citealt{Koopmans2015}) also aim to observe the EoR 21-cm signal. 
 Despite the ongoing efforts, until now we only have a few upper limits on the EoR 21-cm signal (\eg GMRT: \cite{Paciga2011, Paciga2013}; LOFAR: \cite{Yatawatta2013, Patil2017, Gehlot2019, Mertens2020, Mondal2020b}; MWA: \cite{Dillon2014, Jacobs2016, Li2019,Barry2019,Trott2020}; PAPER: \cite{Cheng2018,Kolopanis2019}; HERA: \citealt{hera2021}). 

Foregrounds,  which are three to four orders of magnitude larger than the expected 21-cm signal, are 
the primary challenge for detecting  the redshifted EoR 21-cm signal \citep{Shaver1999, DiMatteo2002,Santos2005, Ali2008,Bernardi2009,Paciga2011,Ghosh2012, Iacobelli2013b, Choudhuri2017a, Choudhuri2020}.  Several methods  have been proposed to separate the 21-cm signal from the foregrounds. One approach, `foreground removal', proposes  to subtract out a foreground model and use the residual data to measure  the 21-cm power spectrum   \citep{Jelic2008, Bowman2009, Paciga2011, Chapman2012, Trott2012,Trott2016}. Recently, foreground removal has been implemented using a  Gaussian Process Regression and this has been used to model and subtract out the  foregrounds from LOFAR \citep{Mertens2018, Mertens2020} and HERA data \citep{Ghosh2020}. The foreground contribution to the cylindrical 21-cm power spectrum $P(k_{\perp}, \, k_{\parallel})$
is predicted to be  confined within a  wedge in the $(k_{\perp}, \, k_{\parallel})$ plane. The  `foreground avoidance' technique  uses the region outside this so called  Foreground Wedge to measure  the 21-cm power spectrum   \citep{Datta2010,Vedantham2012,Thyagarajan2013,Pober2013a, Pober2014,Liu2014a,Liu2014b,Dillon2014,Dillon2015a,Ali2015}. 

It is very difficult to model and subtract out bright point sources located at a considerable angular distance from the phase center. This is due to ionospheric fluctuations and also the inaccuracy of the primary beam pattern at the outer region of the main lobe. The contribution from such sources manifests themselves as oscillatory frequency structures \citep{Ghosh2011a,Ghosh2011b}. Several studies (\eg \citealt{Thyagarajan2015,Pober2016}) have shown that such sources  contaminate the  cylindrical 21-cm power spectrum $P(k_{\perp}, \, k_{\parallel})$ at the higher $k_{\parallel}$ modes. The polarization leakage from any source is also expected to increase with distance from the phase center \citep{Asad2015, Asad2018}.

There are various estimators have been proposed in the literature for the 21-cm power spectrum. Complex visibilities are the primary quantities measured in radio-interferometric observations. Considering image-based estimators 
 (\eg \citealt{Paciga2013}), these have the disadvantage that the estimated power spectrum may be affected by 
 deconvolution errors that arise during image reconstruction. Techniques like the Optimal Mapmaking Formalism \citep{Morales2009} avoid this deconvolution error during imaging. It is advantageous and efficient to estimate the 21-cm power spectrum  from  the measured visibilities directly
 \citep{Bharadwaj2001a, Morales2005, McQuinn2006, Pen2009, Liu2012, Parsons2012b, Liu2014a, Liu2014b, Dillon2015a, Trott2016}.  \citet{Liu2016} presents a visibility-based estimator which incorporates the spherical nature of the sky by 
  using the spherical Fourier-Bessel basis. In addition to the sky signal, the measured visibilities also contain a system noise contribution.  
The noise bias arising from this is an issue for both the image and the visibility-based power spectrum estimators. For example, \citet{Ali2015} have avoided this by using the cross-correlation between the 
 even and odd LST data to estimate the 21-cm power spectrum. However, the full signal available in the data is not used in such an approach. An alternative approach to detect the 21-cm signal \citep{Thyagarajan2018,Thyagarajan2020} uses the fact that the interferometric bispectrum phase is immune to antenna-based calibration errors.  
 
The Tapered Gridded Estimator (TGE; \citealt{Choudhuri2014,Choudhuri2016a,Choudhuri2016b} ) is a novel visibility-based power spectrum estimator which has three distinct advantages. First, the  TGE suppresses the contribution from sources far away from the phase center by tapering the sky response with a tapering window function which falls off faster than the  antenna primary beam (PB). Second, TGE  internally estimates and subtracts out the noise bias and provides an unbiased estimate of the power spectrum. Third, the TGE works with the gridded visibilities, making the estimator 
  computationally fast. The  2D TGE  has been used extensively to study the angular power spectrum  $C_{\ell}$ of the diffuse synchrotron radiation from our Galaxy \citep{Choudhuri2017a,Chakraborty2019a,Chakraborty2019b, Mazumder2020,Choudhuri2020}  and also from within the Kepler supernova remnant \citep{Saha2019}. An  Image-based Tapered Gridded Estimator (ITGE; \citealt{Choudhuri2019})  has been used to measure $C_{\ell}$  of the \HI 21-cm emission from the ISM in different parts of an external galaxy.  
  
  The Multi-frequency Angular Power Spectrum $C_{\ell}(\nu_a, \nu_b)$  (MAPS; \citealt{Datta2007, Mondal2018, Mondal2020a}) can be used to characterize the joint angular and frequency dependence of the sky signal. It is quite straightforward to generalise the 2D TGE for $C_{\ell}$ to a 3D TGE for $C_{\ell}(\nu_a, \nu_b)$ . \citet{Bharadwaj2018} presents a TGE estimator for MAPS, and use this to propose a new technique to estimate the 21-cm power spectrum $P(k_{\perp}, k_{\parallel})$. This estimator proceeds by using the TGE to evaluate the binned MAPS  $C_{\ell}(\Delta \nu)$ where $\Delta \nu= \mid \nu_a-\nu_b \mid$, and then uses this to estimate  $P(k_{\perp}, k_{\parallel})$ through a Fourier transform with respect to $\Delta \nu$. Simulated  $150 \, {\rm MHz}$  GMRT observations have been used to validate this approach, and demonstrate  that  it  is relatively unaffected by missing frequency channels due to flagging. In a recent work \citet{Pal2020} demonstrate this estimator by applying it to   $150 \, {\rm MHz}$  GMRT observations where $47 \%$ of the frequency channels are flagged. It is shown that tapering the sky response considerably suppresses the foreground contribution in the estimated 
  $P(k_{\perp}, k_{\parallel})$. The power spectrum estimates in a rectangular region outside the foreground wedge was found to be relatively free of foreground contamination, and the estimates in this region were spherically binned to  obtain the $2 \, \sigma$ upper limit of $(72.66)^{2}\,\textrm{K}^{2}$ on the mean squared \HI 21-cm brightness temperature fluctuations at $k=1.59\,\textrm{Mpc}^{-1}$.

 In the present paper we consider drift scan observations where  the direction in which the telescope points 
 remains fixed relative  to the earth, however this   changes continuously on the sky due to earth's rotation. 
  This  allows us to achieve the instrumental stability  needed for optimal foreground  characterization across the long observations required  to detect the  faint   cosmological  21-cm signal.    The instrumental   parameters remain unchanged in drift scan observations where  the sky intensity pattern changes continuously with time. Many ongoing EoR radio interferometers, \eg HERA, work solely  in this  mode while this is one of the possible  modes of observation for  the other  interferometers.
  Many different variants of the drift scan technique have been proposed in the literature: 
  the  $m$-mode analysis    \citep{Shaw2014,Shaw2015}  which has been  applied to OVRO-LWA data in   \citet{Eastwood2018};  cross-correlation of the \HI signal in time   (\citealt{Paul2014,Patwa2019} which has be applied to MWA drift scan data \citealt{Patwa2021}), drift and shift method (\citealt{Trott2014}) and fringe-rate  method (\cite{Parsons2016}, as  applied to the  PAPER data). 

 In this paper we  extend  the existing TGE method and propose  the  Tracking 
 Tapered Gridded Estimator (TTGE)   which   uses  drift scan visibility data to estimate the power spectrum. The  sky response of this estimator  tracks a fixed center on the sky, even as different parts of the  sky drift past the telescope's field of view (FoV). 
 The estimated MAPS $C_{\ell}(\nu_a, \nu_b)$  can be used to determine the 21-cm power spectrum $P(k_{\perp}, k_{\parallel})$ using the procedure followed in \citet{Pal2020}. In this paper, we have validated the TTGE using  simulations of drift scan observations  considering the MWA.

 A brief outline of the paper follows. In Section~\ref{sec:drift} we give a brief overview of drift scan observations and setup the notation used in the rest of the paper. In Section~\ref{sec:fsa} we introduce the notion of the flat sky approximation (FSA), and in Section~\ref{sec:ttge} we  use this  to formulate the Tracking Tapered Gridded Estimator (TTGE). In Section~\ref{sec:simulation} we discuss  the simulations that we have  performed,   which is followed by the results in Section~\ref{sec:result}. Finally we summarize and discuss the outcome of this work in Section~\ref{sec:summary}.


\section{Drift scan observations} \label{sec:drift}

The visibility $\vis$ measured at a baseline $\U$ and frequency $\nu$ can be calculated using  

\begin{equation}
\vis = \dBdTnu \int_{UH} d\Omega_{\n}
T\left(\n,\nu \right) A\left(\Delta \n,\nu
\right) e^{2 \pi i \U\cdot \dn },
\label{eq:v1}
\end{equation}
where the antenna primary beam (PB) 
 pattern $A\left(\Delta \hat{\mathbfit{n}},\nu \right)$ quantifies how
 the individual antenna responds to the signal from different
 directions $\hat{\mathbfit{n}}$ on the sky,
 $\m$ refers to the direction in which the antennas are pointing with  
$\Delta \n =\n -\m$, 
  $\dBdTnu = 2 k_B / \lambda^2$ is the conversion factor from
brightness temperature to specific intensity in the Raleigh - Jeans
limit, $T\left(\hat{\mathbfit{n}},\nu \right)$ is the brightness
temperature distribution on the sky and  $d\Omega_{\n}$ is
the elemental solid angle in the direction $\n$. 
For the $d\Omega_{\n}$ integral, it is restricted to the upper hemisphere $(UH)$ .

We consider   a telescope array located   at a latitude $\delta_0$ where the ${\rm (RA,DEC)} =(\alp,\delta_0)$  of the zenith has 
$\delta_0$ fixed  whereas $\alp$  varies   with earth's rotation.
In drift scan observations the baselines $\U$ and the pointing direction $\uv{p}$ all rotate simultaneously   due 
 to earth's rotation, and  we account for this by explicitly showing $\alp$ as an argument $\visa$ for the visibilities.  
 The  telescope array is assumed to have antennas which  are fixed horizontally on the ground (\eg LOFAR, MWA, HERA).
 While it may be possible to electronically steer the pointing direction $\uv{p}$, for simplicity  we consider the situation where $\uv{p}$ point vertically overhead.  It is convenient to introduce   a set of basis vectors  $\uv{e}_1(\alp) ,\uv{e}_2(\alp)$ which are  fixed to the telescope and respectively point towards  East and North  at the location of the array (see Figure~\ref{fig:cs}). 
 We can use these to express the baselines as 
 \begin{equation}
     \U=u \uv{e}_1(\alp) + v \uv{e}_2(\alp) + w \uv{p}\,.
     \label{eq:v2}
 \end{equation}
  We can also express the PB of the telescope in terms of these vectors. For an instrument like MWA,  we model the aperture   as a squares with  sides of length $b$ which are  aligned with the cardinal directions. We can then express  the primary beam pattern as 
\begin{equation}
 A(\dn,\nu)= {\rm sinc}^2 \left( \frac{\pi b \nu \,
   \dn \cdot \uv{e}_1(\alp)}{c} \right) {\rm sinc}^2 \left( \frac{\pi b \nu \,
   \dn  \cdot \uv{e}_2(\alp)}{c} \right)\,.
\label{eq:v7}
\end{equation}
 Figure \ref{fig:cs} shows a schematic  representation of the drift scan observations described here. 
 
 In our simulation of the MWA  visibilities we use   
\begin{equation}
\visa = Q_{\nu} \, \Delta \Omega_{\rm pix} \, \sum _{q=0}^{N_{\rm pix} - 1} \, T(\uv{n}_q,\nu)
\, A(\dn_q,\nu)  e^{2 \pi i \U \cdot \dn_q},
  \label{eq:v8}
\end{equation}
where $\Delta \Omega_{\rm pix}$ refers to the solid angle subtended by
each simulation pixel. The pixels are labelled using $q$,  $\uv{n}_q$
refers to the unit vector towards the $q$-th pixel and  $N_{\rm pix}$ is the number of pixels in the simulation. 
 
\section{Flat sky approximation}\label{sec:fsa}
\begin{figure}
\centering
\includegraphics[scale=0.35]{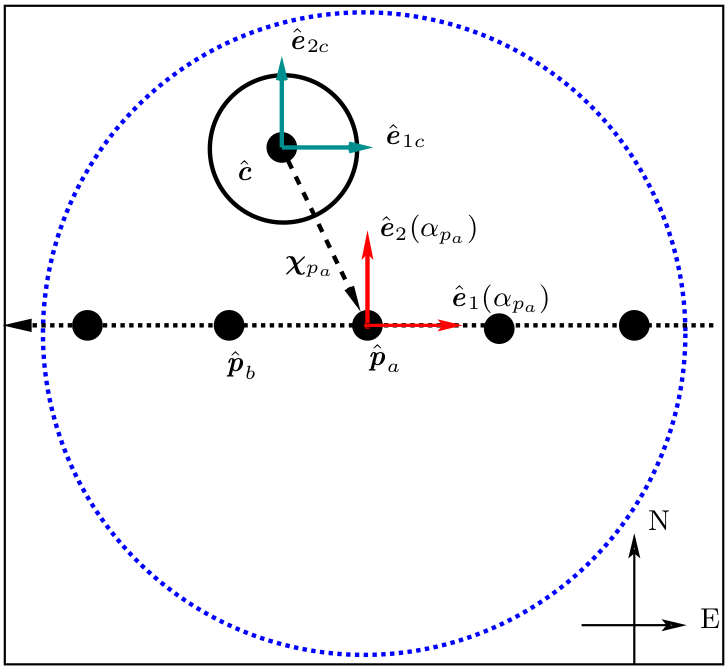}
\caption{This shows a schematic diagram of drift scan observations. Filled circles (marked by 
$\uv{p}_a, \uv{p}_b, \ldots $) are the different pointing direction of the telescope, and the dashed  circle   represents the FWHM  of the MWA primary beam when the telescope is pointing towards $\uv{p}_a$ for which the basis vectors $\uv{e}_1(\alpha_{p_a})$ and $\uv{e}_2(\alpha_{p_a})$ are also shown. 
 $\uv{c}$ refers to  the tracking center for which $\uv{e}_{1c}$ and $\uv{e}_{2c}$ are the corresponding basis vectors. 
 The solid circle  around $\uv{c}$ show the FWHM of the tapering window, and $\bm{\chi}_{p_a}=\uv{p}_a-\uv{c}$.  }
\label{fig:cs}
\end{figure}

It is convenient to develop the Tracking Tapered Gridded  Estimator (TTGE) in the flat sky approximation (FSA) where  the entire analysis is restricted to a small region of the sky centered on a fixed position $(\alc,\delta_c)$ which we refer to as the tracking center. 
 The unit vector $\uv{c}$ points towards the tracking center, and  the unit vectors $\uv{e}_{1c}$ and  $\uv{e}_{2c}$ which are fixed to the sky region of our interest respectively point East and North (see Figure~\ref{fig:cs}).  Considering $\uv{n}$ an arbitrary direction on the sky and  assuming that   $\mid \uv{c} - \uv{n} \mid \ll 1$,   we have 
\begin{equation}
    \uv{n}=\thta \uv{e}_{1c} + \thtb \uv{e}_{2c} + \uv{c} 
    \label{eq:f1}
\end{equation}
where  $\vtht=\thta \uv{e}_{1c} + \thtb \uv{e}_{2c} $ is a 2D vector  on the plane of the sky.  Similarly considering $\uv{p}$ an arbitrary pointing direction of the telescope and assuming  that $\mid \uv{p} - \uv{c} \mid \ll 1 $,  we  have  
 $\uv{p}=\vch + \uv{c}$   where 
\begin{equation}
    \vch=(\alp-\alpha_c) \, \cos \delta_0 \uv{e}_{1c} + (\delta_0-\delta_c) \uv{e}_{2c} \,.
    \label{eq:f2}
\end{equation}
The vector $\vch$ (Figure \ref{fig:cs}) quantifies the angular displacement  between the pointing direction $\uv{p}$ and 
the tracking center $\uv{c}$, whereas $\chi_p=\mid \vch \mid$ quantifies the magnitude of the corresponding angular separation.  
 We use  eqs.~(\ref{eq:f1}) and (\ref{eq:f2})  in eq.~(\ref{eq:v1}) to  express the visibilities as
\begin{equation}
\visa = \dBdTnu \int d^2 \tht \, 
T\left(\vtht,\nu \right) A\left(\vthtp,\nu
\right) e^{2 \pi i \U \cdot (\vthtp) } \,.
\label{eq:f3}
\end{equation}
where throughout the FSA  we only retain terms linear in $\thta,\thtb$, $(\alp-\alpha_c) \, \cos \delta_0 $ and $(\delta_0-\delta_c) $.
Here  we may also consider $\U$ as a 2D vector 
$ \U= u \uv{e}_{1c} + v \uv{e}_{2c}$ whose $(u,v)$ components have the same values as the $(u,v)$ values in eq.~(\ref{eq:v2}).

In the FSA we decompose the sky into Fourier modes $\U$ which are conjugate to $\vtht$.  We then have 
\begin{equation}
    T\left(\vtht,\nu \right)=\int d^2 U \, \tilde{T}(\U,\nu) e^{-2 \pi i \U \cdot \vtht }
    \label{eq:f4}
\end{equation}
where $ \tilde{T}(\U,\nu)$ is the Fourier counterpart of $  T\left(\vtht,\nu \right)$,
and 
\begin{equation}
    A\left(\vtht,\nu \right)=\int d^2 U \, \tilde{a}(\U,\nu) e^{-2 \pi i \U \cdot \vtht }
    \label{eq:f5}
\end{equation}
where $ \tilde{a}(\U,\nu)$, the Fourier transform of $ A\left(\vtht,\nu \right)$ is the antenna aperture power pattern. 

Using these we  express eq.~(\ref{eq:f3}) in Fourier space as  
\begin{eqnarray}
    \visa &=&   e^{-2 \pi i \U \cdot \vch } \, \dBdTnu \,  \int d^2 U^{'} \,  [ e^{2 \pi i (\U -\U^{'}) \cdot \vch } \nonumber \\ 
    & \times & \tilde{a}(\U-\U^{'},\nu) ]\,  \tilde{T}(\U^{'},\nu) \,.
    \label{eq:f6}
\end{eqnarray} 
 The term in the square brackets $[...]$ is the Fourier transform of the shifted primary beam pattern  $ A\left(\vthtp,\nu\right)$.

The sky response of the measured visibilities is determined by $ A\left(\vthtp,\nu\right)$ which is 
centered around  $\vtht=\vch$ which  corresponds to $\uv{p}$ the pointing center of the telescope. 
This keeps on shifting across the sky in drift scan observations. The tracking and tapering  implemented by our estimator  aims to restrict the sky response to a fixed small region of the sky centered around $\vtht=0$ which corresponds to the tracking center $\uv{c}$. This   remains fixed on the sky as different parts of the sky  drift past the zenith.  We  achieve this by multiplying the measured visibilities with a phase factor and then convolving them with a window function $\tilde{w}(\U)$ which is the Fourier transform of the tapering function 
\begin{equation}
    W\left(\vtht \right)=\int d^2 U \, \tilde{w}(\U) e^{-2 \pi i \U \cdot \vtht } \,.
    \label{eq:f7}
\end{equation}

 The tapering function   $W\left(\vtht \right)$ is chosen so that it is peaked around $\vtht=0$ and falls off rapidly away from the center. Here we have used  a Gaussian $ W\left(\vtht \right)=e^{-\theta^2/\theta_w^2}$ which has a full width at half maxima $\theta_{\rm FWHM}=\theta_w/0.6$.  The window function  corresponding  to this is  given by,
\begin{equation}
    \tilde{w}(\U)=\pi \thw^2 \, e^{-\pi^2 U^2 \thw^2 } \,.
    \label{eq:f7b}
\end{equation}

Using eq.~(\ref{eq:f6}),   the convolved visibilities 
\begin{equation}
    \mathcal{V}_c(\U,\nu)=\int d^2 U^{'} \, \tilde{w}(\U-\U^{'}) \, e^{2 \pi i \U^{'} \cdot \vch } \,  \mathcal{V}(\alp,\U^{'},\nu) \, 
    \label{eq:f8}
\end{equation}
can be expressed as 
\begin{equation}
    \mathcal{V}_c(\U,\nu)=\dBdTnu \int d^2 \tht \, 
T\left(\vtht,\nu \right) \, W(\vtht) \, A\left(\vthtp,\nu
\right) e^{2 \pi i \U \cdot \vtht } \,   .
\label{eq:f9}
\end{equation}
where the phase center is shifted to the tracking center, and the function $W(\vtht)$ restricts the sky response to a small region around the tracking center. Note that the tapering window is typically chosen to be considerably narrower than the antenna's primary beam pattern.  

\section{The tracking tapered gridded estimator}\label{sec:ttge}
The idealized calculation presented in the earlier section assumes that the entire $(u,v)$ plane is completely spanned by baselines. In reality we have visibility measurements only at discrete baselines $\U_n$. 
Here we introduce an uniform grid of spacing $\Delta U$ in the $(u,v)$ plane and evaluate the convolved gridded visibilities  on this grid.
Using the label $g$ to denote a grid point with corresponding baseline  $\U_g$ the convolved visibility at this grid point is calculated using 
\begin{equation}
    \vcg(\nu)=\sum_{p}  s_p \sum_{n} \tilde{w}(\U_g-\U_n) e^{2 \pi i \U_n \cdot \vch } \,  \mathcal{V}(\alp,\U_n,\nu) 
\label{eq:c1}
\end{equation}
where the $\int d^2 U^{'}$ integral in eq.~(\ref{eq:f8}) is now replaced by a  sum over different baselines $\U_n$. We now also have an additional  sum over $p$  which combines the measurements  from different pointing direction $\uv{p}$ each with a corresponding weight $s_p$.  The primary beam pattern suppresses the sky signal as  $\chi_p$  increases. The pointing directions $\uv{p}$ with a large value of $\chi_p$  contribute with a lower signal to noise ratio as compared to those with a small value of $\chi_p$.  We can  account for this by suitably choosing  $s_p$. Further,  the sum over $p$ is also restricted to a   range of values  $\chi_p \ll 1$ where the FSA holds. 

Following eq.~(10) of \citet{Choudhuri2016b}  we can write eq.~(\ref{eq:c1}) as 
\begin{equation}
      \vcg(\nu)=Q_{\nu} \int d^2 U \,  \tilde{K}(\U_g,\U,\nu) \,  \tilde{T}(\U,\nu) 
       \label{eq:c3}
\end{equation}
where the kernel $\tilde{K}(\U_g,\U,\nu)$ is defined as 
\begin{eqnarray}
 \tilde{K}(\U_g,\U,\nu)&=&\sum_p s_p  \int d^ 2 U^{'} \,  \tilde{w}(\U_g-\U^{'}) B(\U^{'},\nu)  \nonumber \\      
 &\times& \left[ e^{2 \pi i (\U^{'} -\U) \cdot \vch } \tilde{a}(\U^{'}-\U,\nu) \right]\,.
    \label{eq:c3a}
\end{eqnarray}
Here 
\begin{equation}
    B(\U^{'},\nu) =\sum_n \delta^2(\U^{'}-\U_n)
     \label{eq:c4}
\end{equation}
is the baseline sampling function in the $(u,v)$ plane, note that this will in general vary with frequency if we take  baseline migration into account.

The multi-frequency angular power spectrum (MAPS)  $\cl(\nu_a,\nu_b)$ 
\citep{Datta2007} characterizes the joint frequency and angular dependence 
of the statistical properties of the 
background sky signal. 
We decompose the brightness temperature distribution 
$T_{\rm b} (\n,\,\nu)$ in terms of spherical harmonics
$Y_{\ell}^{m}(\n)$ using 
\begin{equation}
\delta T_{\rm b} (\n,\,\nu)=\sum_{\ell,m} a_{\ell {m}} (\nu) \,
Y_{\ell}^{m}(\n)
\label{eq:alm}
\end{equation}
and define the multi-frequency angular power spectrum (hereafter MAPS) as 
\begin{equation}
\cl(\nu_a, \nu_b) = \big\langle a_{\ell {m}} (\nu_a)\, a^*_{\ell
  {m}} (\nu_b) \big\rangle\, .
\label{eq:cl}
\end{equation}
As discussed in \citet{Mondal2018}, 
we expect  $\cl(\nu_a,\nu_b)$ to entirely quantify the second order statistics of the redshifted  21-cm signal. Assuming the redshifted 21-cm signal  is statistically homogeneous (ergodic) along the line of sight, one can estimate the 3D power spectrum $P(k_{\perp},\,k_{\parallel})$ from $\cl(\nu_a,\nu_b)$ ( \citealt{ Mondal2018, Sarkar2016b}). Then we can recast MAPS as $\cl(\nu_a,\nu_b)=\cl(\Delta \nu)$ where $\Delta \nu=  \mid \nu_b-\nu_a\mid$ \ie the statistical properties of the sky signal depends only on the frequency separation between the channels and not the individual frequencies.  In the flat sky approximation,  the cylindrical power spectrum $P(k_{\perp},\,k_{\parallel})$  of the brightness temperature fluctuations of the ergodic redshifted 21-cm signal  can be expressed as the Fourier transform of $\cl(\Delta \nu)$, and we have \citep{Datta2007}  
\begin{equation}
P(k_{\perp},\,k_{\parallel})= r_{\nu_c}^2\,r_{\nu_c}^{\prime} \int_{-\infty}^{\infty}  d (\Delta \nu) \,
  e^{-i  k_{\parallel} r_{\nu_c}^{\prime} \Delta  \nu}\, \cl(\Delta \nu)
\label{eq:cl_Pk}
\end{equation}
where $k_{\parallel}$ and $k_{\perp}(=\ell/r_{\nu_c})$ are the parallel and perpendicular to the line of sight components of $\bk$ respectively, $r_{\nu_c}$ is the comoving distance corresponding to $\nu_c$ 
the central frequency of our observations and $r_{\nu_c}^{\prime}~(=d r/d \nu)$ is evaluated at $\nu_c$. 
For further details readers are referred to \citet{Bharadwaj2018} and \citet{Pal2020} .

In the FSA the multi-frequency angular power spectrum MAPS  is defined as 
\begin{equation}
    \langle \tilde{T}(\U,\nu_a)  \tilde{T}(\U^{'},\nu_b)  \rangle = \delta^2(\U-\U^{'}) \,  C_{2 \pi U}(\nu_a,\nu_b) \,.
    \label{eq:c5}
\end{equation}
We use $\langle \vcg(\nu_a) \mathcal{V}^*_{cg}(\nu_b) \rangle$ to estimate $C_{\ell}(\nu_a,\nu_b)$ , 
The analysis till now has only considered the sky signal contribution to the measured visibilities. However, 
in addition to the sky signal, every measured visibility also contains  a random noise contribution. 
The noise  in two different visibility measurements ({\it i.e.} different baseline, frequency or time stamp) is   uncorrelated.  Using eq.~(\ref{eq:c1}) and (\ref{eq:c3}),   we then have 
\begin{flalign}
& \langle \vcg(\nu_a) \mathcal{V}^*_{cg}(\nu_b) \rangle = \int d^2 U \, \tilde{K}(\U_g,\U,\nu_a)   \tilde{K^*}(\U_g,\U,\nu_b) & \nonumber \\  
&\times  C_{2 \pi U}(\nu_,\nu_b) +  \delta_{a,b} \sum_{p,n}  \mid s_p  \tilde{w}(\U_g-\U_n) \mid^2 \langle \mid \mathcal{N}(\alp,\U_n,\nu_a) \mid^2 \rangle &  
      \label{eq:c6}
\end{flalign}
where $\delta_{a,b}$ is a Kronecker delta and $\mathcal{N}(\alp,\U_n,\nu_a) $ is the noise contribution to the visibility  $\mathcal{V}(\alp,\U_n,\nu_a) $. 
We approximate this as 
\begin{flalign}
& \langle \vcg(\nu_a)   \mathcal{V}^*_{cg}(\nu_b) \rangle =  \left[ \int d^2 U \,  \tilde{K}_g(\U_g,\U,\nu_a)   \tilde{K^*}_g(\U_g,\U,\nu_b) \right] & \nonumber \\  
& \times C_{\ell_g}(\nu_a,\nu_b) + \delta_{a,b} \sum_{p,n}  \mid s_p  \tilde{w}(\U_g-\U_n) \mid^2 \langle \mid \mathcal{N}(\alp,\U_n,\nu_a) \mid^2 \rangle & .  
      \label{eq:c7}
\end{flalign}
where $\ell_g=2 \pi U_g$ and the integral  in the square bracket $[...]$ is a factor that only depends on the grid point $g$ and the two frequencies $(\nu_a,\nu_b)$. The  kernel $\tilde{K}(\U_g,\U,\nu)$ is peaked around $\U=\U_g$, and it falls off as 
$\mid \U-\U_g \mid$ increases. The approximation used in eq.~(\ref{eq:c7}) holds if $C_{2 \pi U}$ does not vary much over the width of the kernel. In eq.~(\ref{eq:c7})  we see that the correlations between the convolved  visibilities at a grid point $g$ with corresponding baseline $\U_g$ provides an estimate of $C_{\ell_g}(\nu_a,\nu_b)$ at the angular multipole $\ell_g=2 \pi U_g$. There is an additional noise bias when $a=b$ {\it i.e.} the two frequencies are same. This arises from the correlation of a visibility with itself, the noise in two different visibilities being uncorrelated.  

We use this to define the MAPS tracking tapered gridded estimator (TTGE)   
\begin{eqnarray}
\hat{E}_g(\nu_a,\nu_b)&=&M^{-1}_g(\nu_a,\nu_b) {\mathcal Re} \,  [\vcg(\nu_a)   \mathcal{V}^*_{cg}(\nu_b)   \nonumber \\
&-&  \delta_{a,b} \sum_{p,n}  \mid s_p  \tilde{w}(\U_g-\U_n) \mid^2  \mid \mathcal{V}(\alp,\U_n,\nu_a) \mid^2 ] \,. \nonumber \\
\label{eq:c8} 
\end{eqnarray}
Where ${\mathcal Re}()$ denotes the real part, and the normalisation factor $M_g(\nu_a,\nu_b)$ corresponds to  the terms inside the square brackets in eq.~(\ref{eq:c7}). For $a=b$  the second term in the {\it r.h.s.} of eq.~(\ref{eq:c8}) exactly cancels out the noise bias in $  \langle  \mid \vcg(\nu_a)  \mid^2  \rangle $   which arises from the correlation of a visibility with itself.  Some signal  also is subtracted out. However if the number of visibilities which contribute to $\vcg$ be $N_g$ then   $  \langle \mid \vcg(\nu_a) \mid^2 \rangle $ receives contributions from $\sim N_g^2$ visibility pairs whereas the term which cancels out the noise bias receives contributions from only $N_g$ of the self correlation  terms. We  have   $N_g \ll N_g^2$  for very large $N_g$,   and only a very small fraction of the signal  is lost. 

We use all sky simulations to estimate the normalisation factors   $M_g(\nu_a,\nu_b)$.  In any such simulation the signal 
 $T(\uv{n},\nu)$ is a realisation of a  Gaussian random field corresponding to an unit multi-frequency angular power spectrum (UMAPS) for which $\cl(\nu_a,\nu_b)=1$. The simulations consider identical drift scan observations as the actual data, with  exactly the same frequency and baseline coverage. The simulated visibilities $[\V_i(\nu_a)]_{\rm UMAPS}$ are analysed exactly the same as the actual data. We have estimated $M_g(\nu_a,\nu_b)$ using 
 \begin{eqnarray}
&M_g(\nu_a,\nu_b)= \langle  {\mathcal Re} \,  [   \vcg(\nu_a)   \mathcal{V}^*_{cg}(\nu_b) -\delta_{a,b} & \nonumber \\ 
& \sum_{p,n}  \mid s_p  \tilde{w}(\U_g-\U_n) \mid^2  \mid \mathcal{V}(\alp,\U_n,\nu_a) \mid^2]  \rangle_{\rm UMAPS} &
 \label{eq:c9} \,
\end{eqnarray}
 where we have averaged over several random  realisations of the UMAPS  to reduce the statistical uncertainties in the estimated  $M_g(\nu_a,\nu_b)$.  
 
The  MAPS TTGE defined in eq.~(\ref{eq:c8}) 
provides an unbiased estimate of $\cl{_g}(\nu_a,\nu_b)$ at the angular
 multipole $\ell_g=2 \pi U_g$ {\it i.e.} 
\begin{equation}
\langle {\hat E}_g(\nu_a,\nu_b) \rangle = \cl{_g}(\nu_a,\nu_b)
\label{eq:a5}
\end{equation}
 We  use this to define the binned Tracking Tapered Gridded Estimator for bin $a$ 
\begin{equation}
{\hat E}_G[a](\nu_a,\nu_b) = \frac{\sum_g w_g  {\hat E}_g(\nu_a,\nu_b)}
{\sum_g w_g } \,.
\label{eq:a6}
\end{equation}
where $w_g$ refers to the weight assigned to the contribution from any particular 
grid point $g$. For the analysis presented in this paper we have used the 
weight $w_g=M_g(\nu_a,\nu_b)$  which roughly averages the visibility correlation 
$\V_{cg}(\nu_a) \,  \V_{cg}^{*}(\nu_b)$ across  all the  grid points
 which are sampled by the baseline distribution. 
The binned estimator  has an expectation value 
\begin{equation}
\bar{C}_{\bar{\ell}_a}(\nu_a,\nu_b)
  = \frac{ \sum_g w_g \cl{_g}(\nu_a,\nu_b)}{ \sum_g w_g}
\label{eq:a7}
\end{equation}
where $ \bar{C}_{\bar{\ell}_a}(\nu_a,\nu_b)$ is the bin averaged
  multi-frequency angular  power spectrum  (MAPS) at 
 \begin{equation}
\bar{\ell}_a =
\frac{ \sum_g w_g \ell_g}{ \sum_g w_g}
\label{eq:a8}
\end{equation}
which is the   effective angular multipole  for bin $a$.

As mentioned earlier, the window function $W(\vtht)$ restricts the sky response to a small region around the tracking center $\uv{c}$, and the estimated $C_{\ell(\nu_a,\nu_b)}$ is only sensitive to the signal from this  small region of the sky. However, the telescope's primary beam pattern covers a substantially larger region of the sky (Figure \ref{fig:cs}), and the sky coverage is  enhanced even further as the pointing direction sweeps across the sky in a drift scan observation.  The question is ``How to utilise the entire sky coverage of the observations?".  Here we proceed by considering a set of $N_{TC}$ different tracking centers  $\uv{c}_1, \uv{c}_2,\uv{c}_3,...$ which span the  sky region covered in the drift scan observations (Figure \ref{fig:centers}).  We use $[C_{\ell(\nu_a,\nu_b)}]_{c_1},  [C_{\ell(\nu_a,\nu_b)}]_{c_2}, [C_{\ell(\nu_a,\nu_b)}]_{c_3},...$ to denote the  estimated bin-averaged MAPS corresponding to $\uv{c}_1, \uv{c}_2,\uv{c}_3,...$ respectively. The average 
\begin{equation}
    [C_{\ell(\nu_a,\nu_b)}]_{{\rm avg}}=(N_{TC})^{-1} \, \sum_{a=1}^{N_{TC}}  [C_{\ell(\nu_a,\nu_b)}]_{c_a}
\label{eq:tc_avg}
\end{equation}
combines the signal from different parts of the sky. The $[C_{\ell(\nu_a,\nu_b)}]_{c_a}$ estimated at different tracking centers  may be considered to be independent provided that  the   tracking centers are sufficiently far apart so that the overlap between
the respective sky coverage may be  neglected. We then expect the cosmic variance to be reduced by a factor $\sqrt{N_{TC}}$ with respect to that for a single tracking center.

\section{Simulations}\label{sec:simulation}
\begin{figure}
    \hspace*{-0.95cm} 
    \includegraphics[scale=0.45]{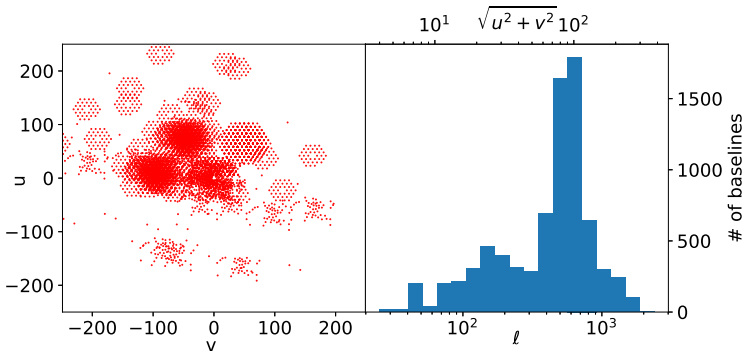}
    \caption{The left panel shows the MWA baseline $(uv)$ distribution in wavelength units.  Dividing the baselines into 20 logarithmic bins of equal spacing, 
    the  right panel shows the number of baselines in each  $\ell$ bin. }
    \label{fig:bl}
\end{figure}
In this section we give an overview of the all-sky simulations that we have carried out to validate 
the TTGE. The  entire analysis is restricted to a single frequency  
$\nu_c = 154 \, {\rm MHz}$ with corresponding wavelength $\lambda_c=1.95 \, {\rm m}$, and we have not explicitly shown the frequency dependence in much of the subsequent discussion.  
We assume that the sky signal $T(\n)$ is a Gaussian random 
field with an angular power spectrum $\cl$ which we have modelled as a power law 
\begin{equation}
\cl^M = A \, \left(\frac{\ell}{\ell_0}\right)^n \,,
\label{eq:cl_1}
\end{equation}
arbitrarily normalized  to $A = 10 \, {\rm mK}^2$ at $\ell =
\ell_0=1$.  We have considered the power law index to have value 
$n=-1$ in our analysis.
 
\begin{figure}
    \centering
    \includegraphics[trim={1cm 0 0 0},clip,scale=0.50]{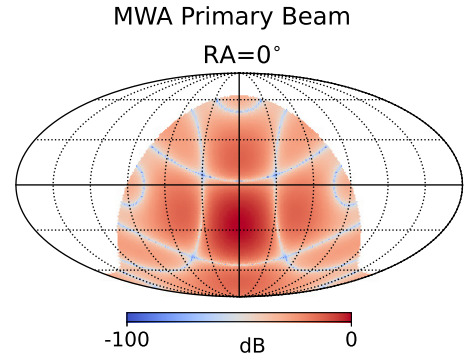}
    \caption{This shows our model for the MWA primary beam pattern in the upper half (UH) of the sky. The bean is pointing at zenith, \ie at DEC=$-26.7^{\circ}$ and at RA=$0^{\circ}$.}
    \label{fig:beam}
\end{figure}
The sky signal was simulated using the package \texttt{HEALPix} \citep[Hierarchical Equal Area isoLatitude Pixelization of a sphere;][]{Gorski2005}.  The left panel of Figure~\ref{fig:bl} shows the MWA baseline distribution used for the simulations. We note that 
 $99$ percent of the baselines are smaller than 500 m  which  corresponds to an angular multipole  $\ell \approx 2 \pi \times 500 \, {\rm m}/\lambda_c=1570$ (see right panel of Figure ~\ref{fig:bl}). In the simulations, we have  set $\ell_{max}= 1535$ which corresponds to  a  pixel size of $6.87^{'}$.

\begin{figure}
    \centering
    \includegraphics[scale=0.36, angle = 0]{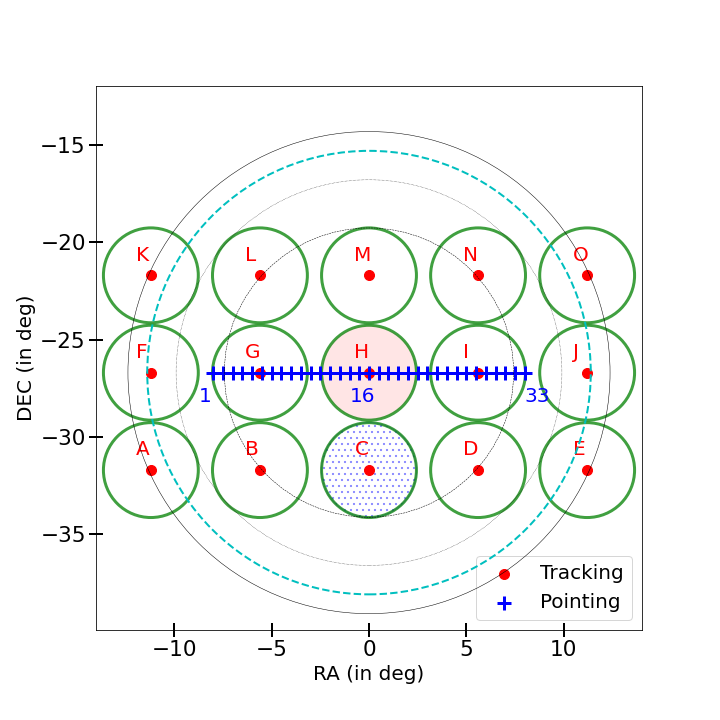}
    \caption{This shows the pointing and tracking centers used in the simulations. Blue points (shown with `+') show the drift scan track \ie the pointing centers (PC) and the red filled circles show the tracking centers (TC). Cyan dashed-circle here shows the FWHM $(\sim 23^{\circ})$ of the MWA primary beam with PC  at $(0^{\circ}, -26.7^{\circ})$. Large thin black circles show the FWHM $\sim 15^{\circ}, 20^{\circ} \, {\rm and }\, 25^{\circ}$ of the wide tapering windows corresponding to the $\theta_w=9^{\circ}, 12^{\circ} \, {\rm and }\, 15^{\circ}$. Whereas, the smaller green circles show the FWHM $(\sim 5^{\circ})$ of the tapering window corresponding to $\theta_w=3^{\circ}$. The two TC for which the results have been shown later are  highlighted in different fill styles. } 

    \label{fig:centers}
\end{figure}

 We have used  eq.~(\ref{eq:v8}) to simulate  the visibilities expected in drift scan observations with   the MWA. The MWA primary beam pattern  $A(\dn)$ was modeled using eq.~(\ref{eq:v7}) with $b = 4 {\rm m} $ (see Figure~\ref{fig:beam}). The MWA  array is  located at a latitude of   $-26.7^{\circ}$. We have considered drift scan observations where the antennas point vertically overhead, whereby the DEC for the pointing direction is fixed at $\delta_0=-26.7^{\circ}$. The RA of the pointing direction changes as it drifts across the sky. Here we have considered an observation where  $\alp$ the RA of the  pointing direction  varies  from $-8.0^{\circ}$ to $8.0^{\circ}$.  The  visibilities are typically recorded  at  an RA interval of $\Delta \alpha_p=7.5''$ in drift scan observations at the MWA \citep{Patwa2021}. This would require us to simulate visibilities for $361$ different pointing directions. However, in order to reduce the volume of computation we have considered  a larger interval  $\Delta \alpha_p =0.5^{\circ}$ whereby we  simulate the visibilities only along $33$ pointing centers (PC) which are labelled as PC$=1,2,3,..., 33$ as shown in Figure~\ref{fig:centers}.  The different parameters values used  for the simulation are summarized  in Table~\ref{tab:sim_par}.

\begin{table*}
\begin{center}
{\renewcommand{\arraystretch}{1.5}%
\begin{tabular}{||l|c||l|c||}
 \hline
 Wavelength $(\lambda_c)$& 1.94 m & Frequency $(\nu_c)$ & 154.24 MHz  \\
\hline
 FWHM & $\sim 23^{\circ}$ & Antenna dimension $(b)$  & 4 m   \\
 \hline
 Pointing center RA $(\alpha_p)$& $[-8.0^{\circ}, -7.5^{\circ}, \ldots , \, 8.0^{\circ}]$ & Pointing center DEC $(\delta_p)$ & $-26.7^{\circ}$  \\
 \hline
 Tracking center RA$(\alpha_c)$& -11.2, -5.6, 0.0, 5.6,  11.2 & Tracking center DEC $(\delta_c)$ & -31.7, -26.7,  -21.7  \\
 \hline
\end{tabular}}

\end{center}
\caption{List of different simulation parameters.}
\label{tab:sim_par}
\end{table*}
 To incorporate noise in our simulated visibilities, we have first estimated $\sigma_s$ the standard deviation of the simulated signal visibilities. We construct visibilities with noise at the central frequency using 
\begin{equation}
    \visnonu = \signonu + \nsnonu
    \label{eq:ns1}
\end{equation}
where, $\signonu$ is the signal visibility and $\nsnonu = a + i \, b$. Here $a \, {\rm and} \, b$ are Gaussian random variables with zero mean and standard deviation $\sigma_n=n \sigma_s$ where $n$ characterizes  the strength of the noise present in the simulated  visibilities. In this paper we have used $n=2 $ and 5 for demonstration purpose.

\section{Results} \label{sec:result}
In this section we discuss the results obtained from  applying the TTGE  to simulated    MWA visibilities. Here we consider a single frequency channel to validate our formalism for the TTGE, however it will be straightforward to generalize for multi-frequency observations. We have used 50 realizations of the model angular power spectrum eq.~(\ref{eq:cl_1}) to generate the visibilities for validation.  These were used to estimate  the mean $\cl$ and  its  1$\sigma$ rms  fluctuations. It may also be noted that 
we have used  50 realizations of the unit angular power spectrum (UMAPS) to estimate the normalization constant $M_g$ (eq.~\ref{eq:c9}). Further, for the present  analysis,   the visibilities with baselines $U$ in the range $2 \leq U \leq 250$ (\ie multiplole range $15 \leq \ell \leq 1535$) were divided in 20 equally spaced logarithmic bins.

  The TTGE offers a wide variety of possibilities  for analyzing the drift-scan  MWA data. Here we have considered three of these possibilities in order to demonstrate and validate the capabilities of TTGE.  We first consider a wide tapering function  whose FWHM is comparable to the FWHM of the MWA primary beam. One would use such a tapering to 
 mainly  suppress the foreground contribution from the side lobes of the primary beam. In contrast, in some situations it may be desirable   to consider a narrow tapering function in order to mitigate  the contribution from some foreground sources which are located very close to or even within the main lobe of the primary beam. We have separately considered these possibilities in sub-sections \ref{sub:1} and \ref{sub:2} respectively. In both cases, we have coherently combined the signal from different pointing directions to estimate $C_{\ell}$ for  a single tracking center (TC) which is labelled TC = H in Figure~\ref{fig:centers}. The narrow window function uses only a small portion  of  MWA's  field of view to estimate $C_{\ell}$  for a single TC.  However, we can overcome this limitation to some extent by considering  a set of TCs which span the full  angular extent of the observations. We use  TTGE to separately estimate $C_{\ell}$  for each of these TCs, and then incoherently combine these estimates (eq.~\ref{eq:tc_avg}).  We have demonstrated this in sub-section \ref{sub:3} where we consider $15$ different TCs labelled TC=A,B,C,..., N,O as shown in Figure~\ref{fig:centers}.  The sky coordinates of the different TC are given  in Table~\ref{tab:sim_par}.

\begin{figure*}
    \hspace{-1.0cm}
    \includegraphics[scale=0.29]{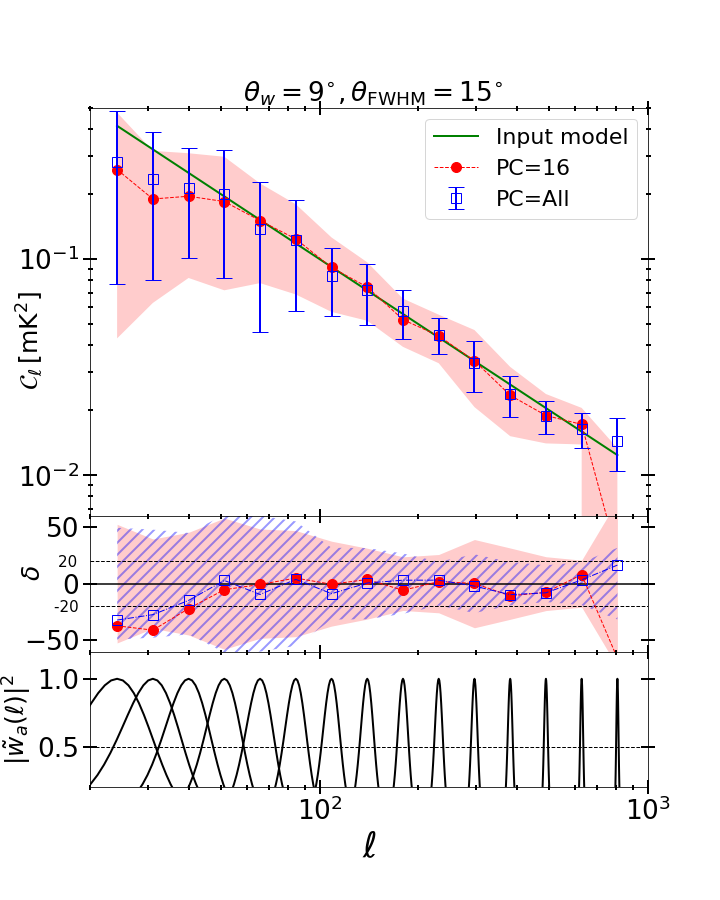}
    \hspace{-1.0cm}
    \includegraphics[trim={3cm 0 0 0},clip, scale=0.29]{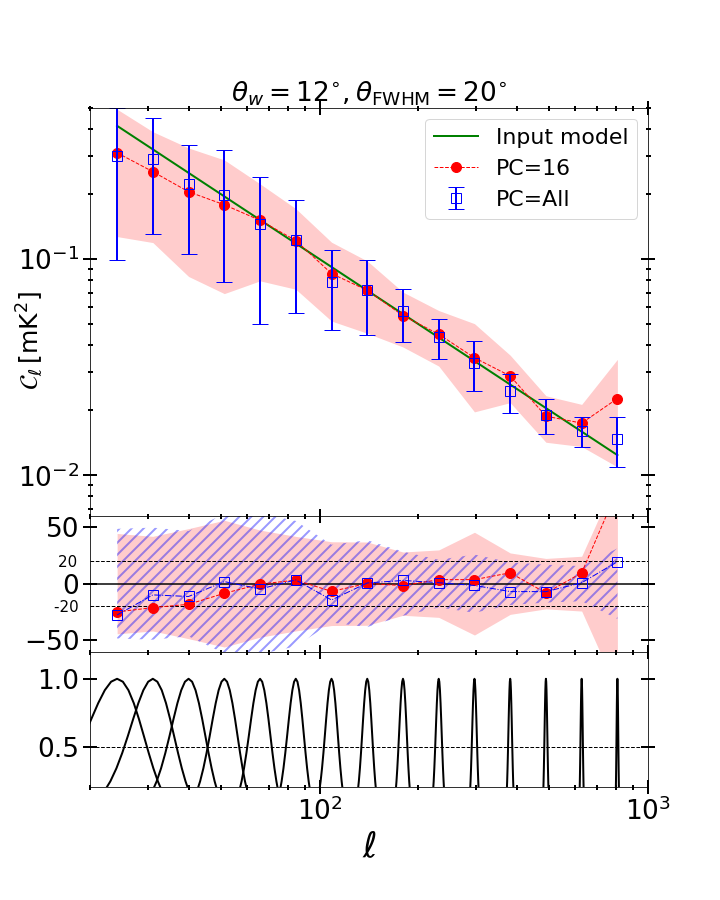}
    \hspace{-1.0cm}
    \includegraphics[trim={3cm 0 1.5cm 0},clip,scale=0.29]{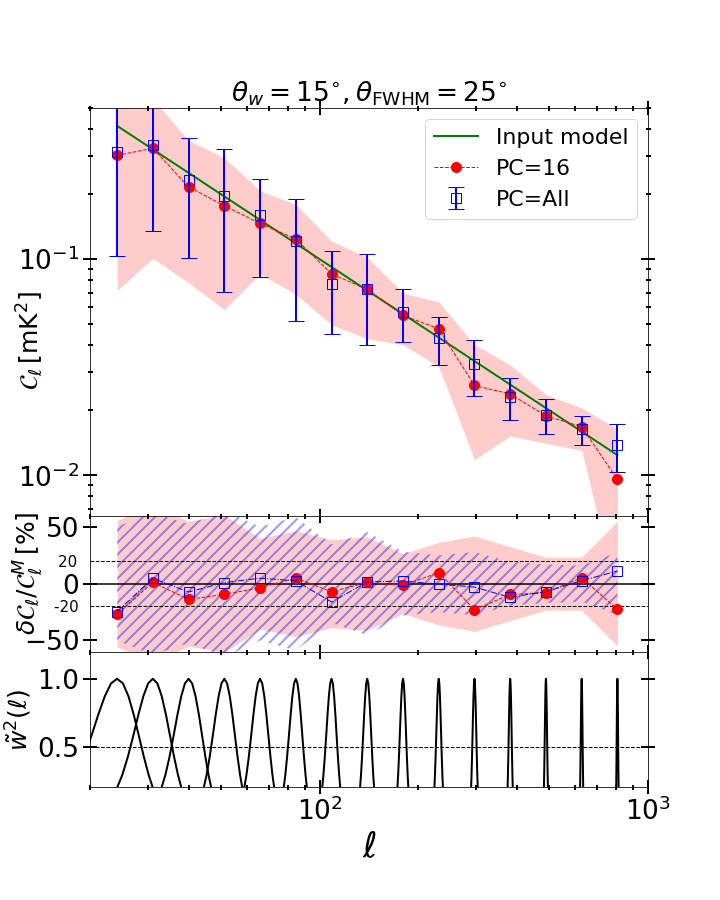}
  
    \caption{Upper sub-panels of the figure show $\cl$ estimated for TC=H using TTGE with $\thf=15^{\circ}, 20^{\circ} \, {\rm and} \, 25^{\circ}$.  The solid line shows the input model (eq.~\ref{eq:cl_1}) used for the simulations which have a noise level of $2 \sigma_s$.  50 realizations of the simulations were used to estimate the mean and $1 \sigma$ errors shown here.  The  red circles and the shaded region show the results for PC=16 whereas the blue squares and error-bars correspond to PC=ALL. Considering the  $\cl$ estimates in the upper sub-panel,  the middle sub-panel shows the corresponding percentage deviation  relative to  the input model, the shaded and hatched regions show the $1\sigma$  statistical fluctuations for PC=16 and PC=ALL respectively.  Lower sub-panels show the variation of normalized window function $|\tilde{w}_a^2(\ell)|^2 = |\tilde{w}((2 \pi)^{-1} [\ell  - \bar{\ell}_a])/\pi \thw^2|^2$, where $\bar{\ell}_a$ is the effective angular multipole of the $a$-th $\ell$ bin.  This variation shows how the measurements in the adjacent bins are correlated due to the overlap of the window functions.}

    \label{fig:fwide}
\end{figure*}

\begin{figure*}
    \hspace{-1.0cm}
    \includegraphics[scale=0.29]{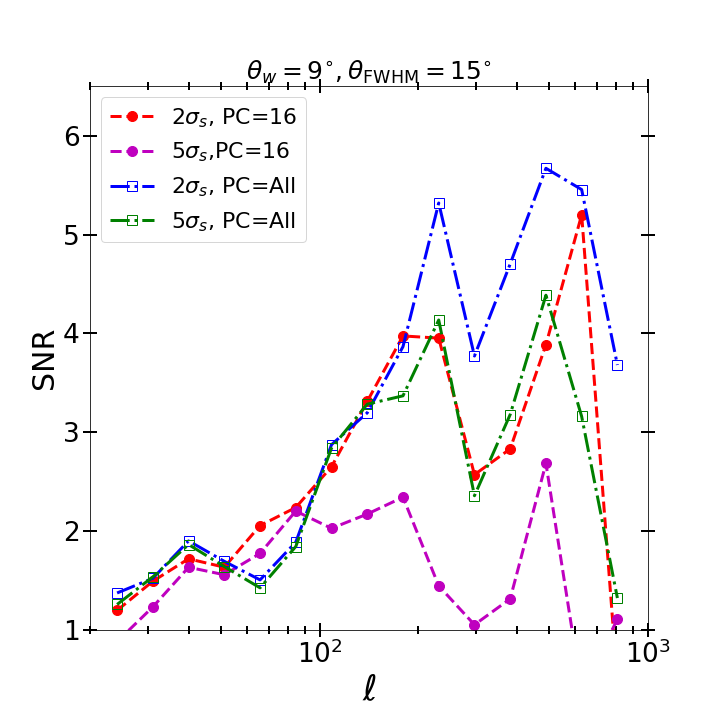}
    \hspace{-1.0cm}
    \includegraphics[trim={3cm 0 0 0},clip, scale=0.29]{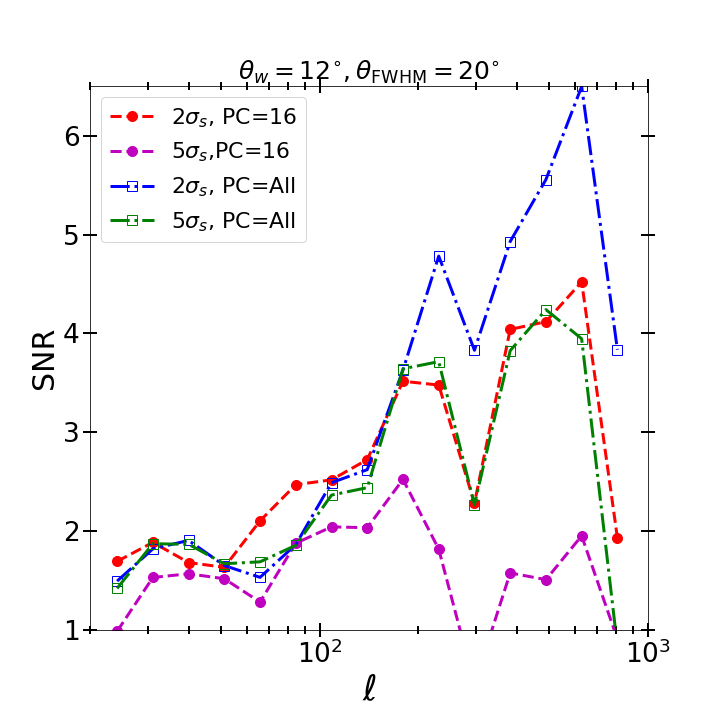}
    \hspace{-1.0cm}
    \includegraphics[trim={3cm 0 1.5cm 0},clip,scale=0.29]{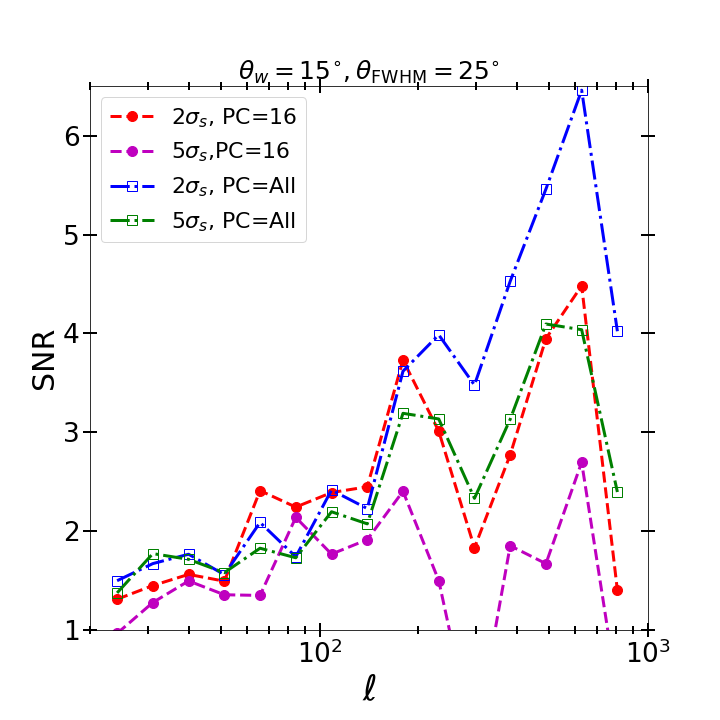}
    \caption{Left, middle and right panels show a comparison of the signal-to-noise-ratio (SNR) for $\thf=15^{\circ}, 20^{\circ} \, {\rm and} \, 25^{\circ}$ respectively. In addition to the $2 \sigma_s$ noise level shown in Figure~\ref{fig:fwide}, here we have also considered a noise level of $5 \sigma_s$.}

    \label{fig:fwide_snr}
\end{figure*}

 \subsection{Wide tapering function}
 \label{sub:1}
 Here we have considered three different values of $\thw$ namely $\thw = 9^{\circ}, 12^{\circ} \, {\rm and} \,15^{\circ}$ which respectively correspond to $\theta_{\rm FWHM}= 15^{\circ}, 20^{\circ} \, {\rm and} \, 25^{\circ}$ 
for the tapering function. The three large black solid circles in Figure~\ref{fig:centers} respectively indicate these  $\theta_{\rm FWHM}$ values, whereas the  cyan dashed circle indicates $\theta_{\rm FWHM}=23^{\circ}$ which corresponds to the main lobe of the MWA primary beam. The results, which we discuss below, are presented in Figure~\ref{fig:fwide}.  These results are all for a single tracking center (TC= H), and the  simulated visibilities have a noise contribution with $\sigma_n=n  \sigma_s$ where $n=2$. 

The  right panel of Figure~\ref{fig:fwide} shows the results for $\theta_{\rm FWHM}=25^{\circ}$ which is slightly larger than  the FWHM of the MWA primary beam pattern (Figure~\ref{fig:centers}). As mentioned earlier, in this case tapering serves to only suppress the side-lobe contribution.  The upper sub-panel shows the estimated $\cl$ and 1$\sigma$  errors  for a single pointing center (PC =16) and  when all the pointing centers (PC=ALL)  are combined with equal weights $s_p=1$ (eq.~\ref{eq:c1}). The 1$\sigma$ error for PC=16 and PC=ALL  are shown by the red shaded region and the blue error bars respectively. In both cases we find that the estimated $\cl$ is in good agreement with the input model $\cl^M$ over nearly the entire $\ell$ range.  Further, the system noise contributions in the different PC are independent, and we expect smaller $1\sigma$ error-bars   for PC=ALL in comparison to PC=16. We see that this is borne out in the results shown here. The middle sub-panels show the percentage  deviation $\delta  =100 \, \% \times \, (\cl -\cl^M)/ \cl^M)$,  and  the dashed horizontal lines mark  $\delta   =   \pm 20 \%$. The red shaded region and blue hatched region show the $\sigma/\cl^M$ for PC=16 and PC=ALL respectively. Considering PC=16, we find that  $\delta   \le \pm 20\%$  in the range $30 < \ell \leq 800$.  The larger deviation at small $\ell$ is due to the convolution with the kernel $\tilde{K}(\U_g,\U,\nu)$ (eq.~\ref{eq:c3}) which becomes important at small baselines. The behaviour of the kernel is largely dominated by $\tilde{w}(\U_g-\U^{'})  $ (eq.~\ref{eq:c3a}), and we expect the convolution to become important at  $\ell < 180^{\circ}/\theta_{\rm FWHM} \sim 7.2$. The value of $\delta$ exceeds $20 \, \%$  at large $\ell$ ($ > 800$)  because the magnitude of $\cl^M$ becomes small and approaches that  of   the fluctuations due to noise.  The results for  PC=ALL are very similar to those for PC=16, with the difference that   $\delta < 20 \, \%$ even at large $\ell$ ($ > 800$). This is because the noise level is reduced when we combine different PC.  This is also reflected in the error-bars which  are reduced to some extent at large $\ell$  ($\ge 200$).  Ideally, we expect the errors to be reduced by a factor of $33$ if these are entirely due to the system noise. However,  the errors at small $\ell$ are cosmic variance dominated and we do not find any reduction here when we combine different PC. We do find some reduction at large $\ell$, however this is considerably smaller that $33$ due to the cosmic variance contribution.   The lower sub-panel shows the normalized  window function  $|\tilde{w}_a^2(\ell)|^2 = |\tilde{w}((2 \pi)^{-1} [\ell  - \bar{\ell}_a])/\pi \thw^2|^2$ where $\bar{\ell}_a$ is the effective angular multipole of the $a$-th $\ell$ bin. The overlap of  $|\tilde{w}_a^2(\ell)|^2$ between any two bins provides an estimate of the correlation between the  $\cl$ values estimated in these two bins. We see that while there is significant overlap between the two lowest $\ell$ bins, this  falls  below $|\tilde{w}_a^2(\ell)|^2 \sim 0.5$ at larger $\ell$   and   there is practically no overlap between the adjacent bins for  $\ell > 60$ (fourth bin onward).  This indicates that the values of $\cl$ are largely uncorrelated the fourth bin onward.

The  middle and the left panels of Figures~\ref{fig:fwide} show the results for the tapering function with $\thf = 20^{\circ} \, {\rm and } \, 15^{\circ}$. These results are very similar to the results in right panel and we expect the estimated $C_{\ell}$ to deviate further from $C_{\ell}^M$ at the small values of $\ell$ as compared to $\thf=25^{\circ}$. For  $\theta_{\rm FWHM}=20^{\circ}$, we find that the estimated $C_{\ell}$ is in good agreement with $C_{\ell}^M$ in the range $30 < \ell \leq 800$. The $\ell$ range over which the deviation $\delta$  remain within  $\approx \pm 20\%$ does not differ much when compared with $\thf = 25^{\circ}$ except for the fact that the deviation from the $C_{\ell}^M$ in the small $\ell$ bins increases (middle sub-panel). Considering  PC=ALL, we see that the behaviour of the estimated $\cl$ is quite similar to that for a single pointing with a  small but  noticeable deterioration in  $\delta$ at low-$\ell$. Considering these small bins, we find that the overlap between the adjacent window functions increase  (shown in lower sub-panel) and this possibly makes the measurements in these $\ell$ bins erroneous.  Overall, the results for $\thf=20^{\circ}$ are all qualitatively similar to those for  $\thf=25^{\circ}$, with the difference that the values of  $\delta$  increase to some extent for  $\thf=20^{\circ}$. 
For the smaller tapering function where $\thf=15^{\circ}$ (left panel), the results are qualitatively similar to that of $\thf=20^{\circ} \, {\rm and} \, 25^{\circ}$ except at first three small $\ell$ bins. In these small values of $\ell$, we find the deviation between estimated $C_{\ell}$  and $C_{\ell}^M$ successively increases with decrease in $\thf$. This is primarily due to the fact that a narrower tapering function results in higher cosmic variance and wider window function (shown in lower sub-panel). As a consequence the errors in the first few small $\ell$ bins are increased.

 We consider  the signal-to-noise ratio (SNR) to quantify the improvement in the errors when we  combine the  data from different PC. Right panel of Figure~\ref{fig:fwide_snr} show a comparison between the SNR achieved considering  PC=16 and PC=ALL for $\thf = 25^{\circ}$. Considering PC=16 with $2 \sigma_s $ noise, we find  a maximum  SNR of $4.3$ and the SNR $>2$ for a large number of $\ell$ bins in the range $100 < \ell < 800$. Increasing the noise level to  $5 \sigma_s$, we have a maximum SNR of $2.7$ {\it i.e.} a factor of $1.6$ smaller.  The  degradation   in SNR is not uniform across all $\ell$. This is more noticeable at large $\ell$ which are system noise dominated,  whereas the  SNR does not decrease much at small $\ell$ which are cosmic variance dominated. 
 The SNR  values  increase moderately  when we consider PC=ALL.  We now have  a maximum SNR  of  $6.5 $  and $4$ for  $2 \sigma_s$ and $5 \sigma_s$ noise respectively.  Comparing PC=16 with PC=ALL,  we see that for $\ell < 90$ the SNR values do not improve much  primarily because  the errors are cosmic variance dominated. There is a relatively larger increase in SNR at larger $\ell$ which are system noise dominated.  In all cases we also notice a sudden dip in the SNR around  $250 < \ell < 350$. This can be attributed to  the small number of baselines in this $\ell$ range  compared to the other $\ell$ bins (right panel of Figure~\ref{fig:bl}).

Middle panel of Figure~\ref{fig:fwide_snr} shows the  SNR for $\thf = 20^{\circ}$ whereas the left panel shows the results for $\theta_{\rm FWHM}=15^{\circ}$. The solid angle subtended by the tapering function  scales  as $\thf^2$ as $\thf$ is decreased from $25^{\circ}$ to $20^{\circ}$. Considering an uniform baseline distribution, we expect the cosmic variance to increase by a factor $(1.25)^2$ or the  SNR is expected to decrease by a factor of $1.5$ provided it is entirely cosmic variance dominated. Comparing the SNR values we do not see much decrement in comparison to  $\thf=25^{\circ}$ (right panel).  Similarly we find that in the small  $\ell$ range  $< 90$, the SNR is decreased . However,  we find that the decrease in the SNR is quite small ($\lesssim 10 \%$)  in the $\ell$ range  $< 90$ which is expected to be cosmic variance dominated. Considering the SNR for the smaller window function the results are qualitatively similar to that of $\thf=20^{\circ} \, {\rm and} \, 25^{\circ}$ (shown in left panel). Similar to the middle panel, we do not notice any substantial degradation in the SNR with decreasing $\thf$.  We interpret this behaviour of the SNR as arising due to the  the non-uniform baseline distribution. Rather than the solid angle on the sky,  we see that the baseline density plays   a more important role in determining the behaviour of the SNR in the individual $\ell$ bins. 
  
It is worth noting that the primary goal of the TTGE is to provide an unbiased estimate of the sky signal while suppressing unwanted contamination far away from the TC. Here we have demonstrated that given an MWA like telescope the TTGE can provide an unbiased estimate of the sky signal. In Appendix~\ref{sec:app1}, we demonstrate the ability of suppressing foreground contamination from far away of the TC. There, we also compare the degree of suppression with a typical visibility correlation estimator.

\begin{figure*}
    \hspace{-1.0cm}
    \includegraphics[scale=0.29]{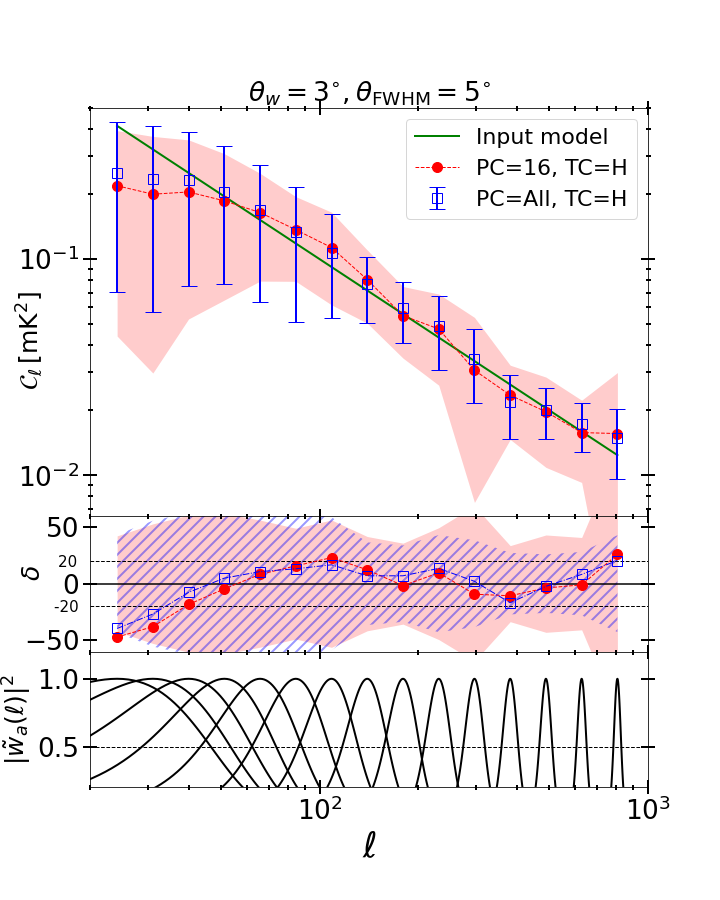}
    \hspace{-1.0cm}
    \includegraphics[trim={3cm 0 0 0},clip, scale=0.29]{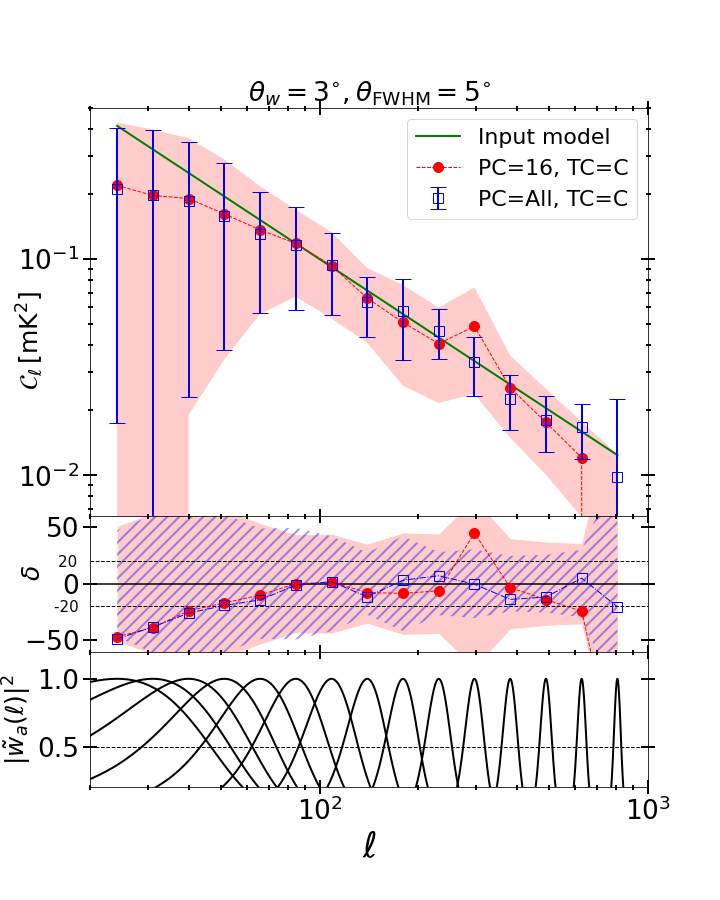}
    \hspace{-1.0cm}
    \includegraphics[trim={3cm 0 1.5cm 0},clip,scale=0.29]{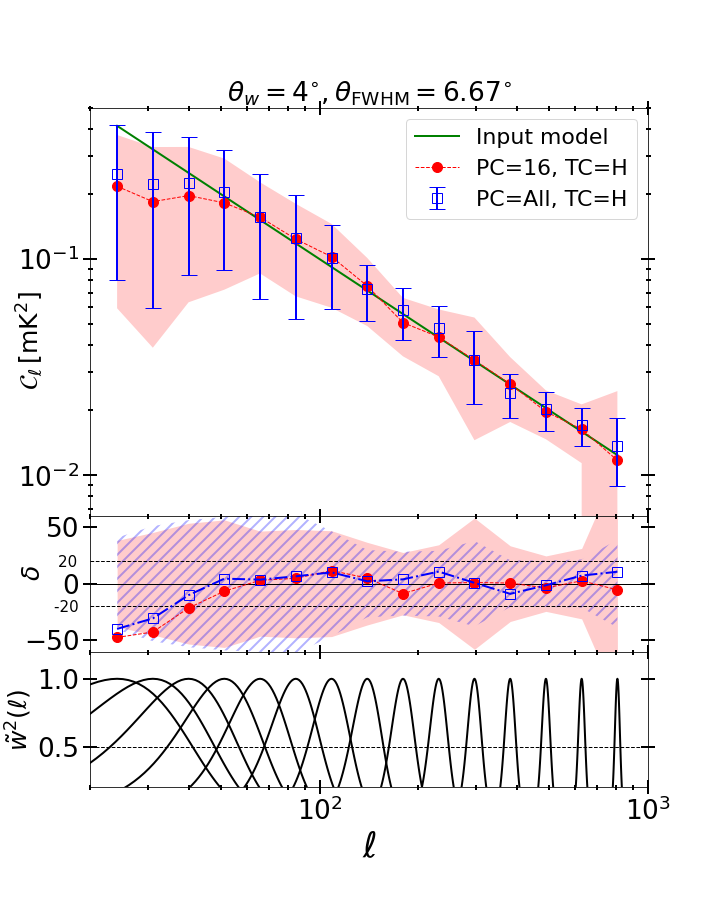}
    \caption{Same as Figure~\ref{fig:fwide} except the $\thf =5^{\circ} \, {\rm and} \, 6.67^{\circ}$. Here we also show results for TC=C.}
    \label{fig:f1}
\end{figure*}
\begin{figure*}
    \hspace{-1.0cm}
    \includegraphics[scale=0.29]{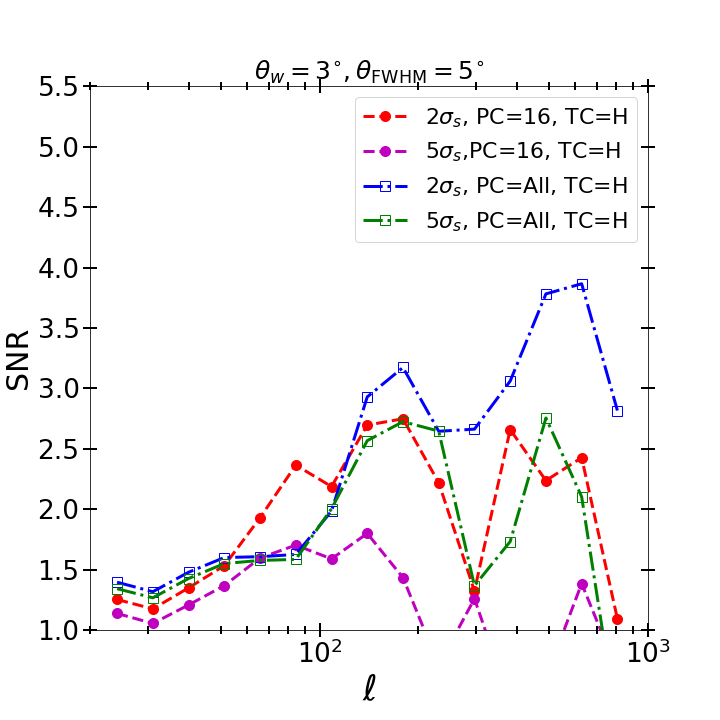}
    \hspace{-1.0cm}
    \includegraphics[trim={3cm 0 0 0},clip, scale=0.29]{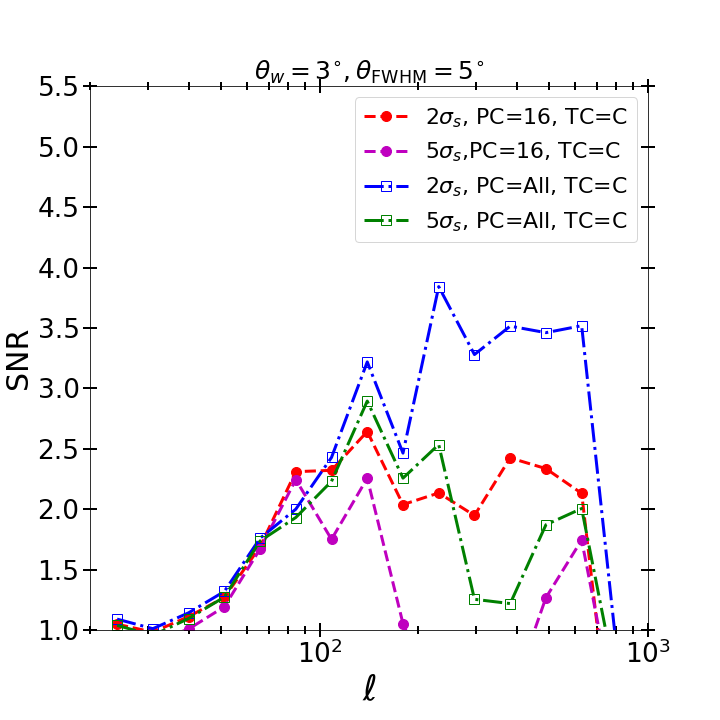}
    \hspace{-1.0cm}
    \includegraphics[trim={3cm 0 1.5cm 0},clip,scale=0.29]{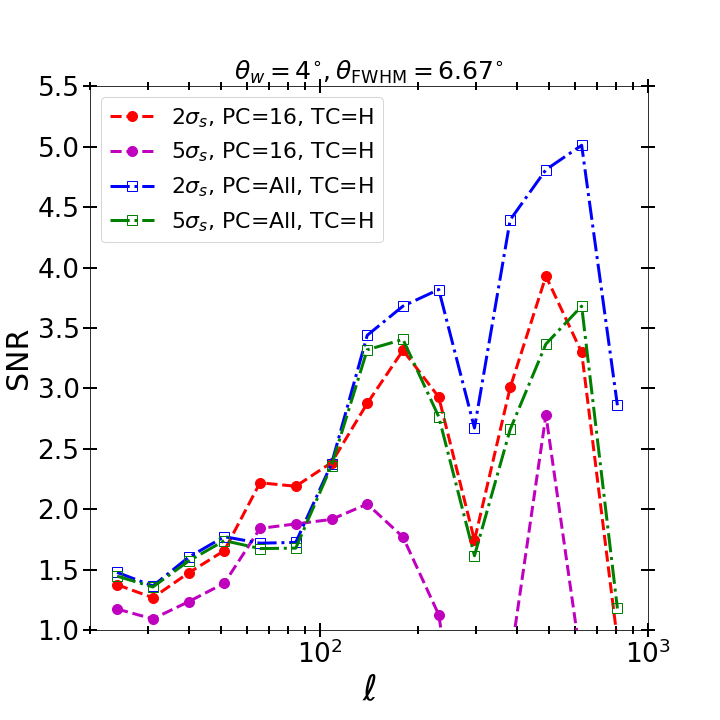}
    \caption{Same as Figure~\ref{fig:fwide_snr} except the $\thf =5^{\circ} \, {\rm and} \, 6.67^{\circ}$. Here we also show results for TC=C.}

    \label{fig:f1_snr}
\end{figure*}

\subsection{ Narrow tapering function}\label{subsec:cp}
\label{sub:2}
Next we consider a situation where it is necessary to mitigate the contribution from foreground sources that are located very close to or even within the main lobe of the primary beam. one may use small tapering functions and strategically place them within the  primary lobe of the telescope beam to minimize the foreground contamination. To demonstrate this situation we have placed fifteen TCs with small tapering functions within the primary lobe of the telescope beam (green circles in Figure~\ref{fig:centers}). 
Here of the width of the tapering function is $\thw = 3^{\circ}$ which corresponds to $\thf=5^{\circ}$. This is significantly narrow compared to the width of the tapering functions discussed earlier (see Section~\ref{sub:1}). Considering narrow tapering functions, the results (Figure~\ref{fig:f1}) and the SNRs (Figure~\ref{fig:f1_snr}) are shown in very similar form as described earlier in Figure~\ref{fig:fwide} and ~\ref{fig:fwide_snr}  (see Section~\ref{sub:1} for detailed description).

First we consider a situation where TC = H (shaded circle in Figure~\ref{fig:centers}), and the left panel of Figure~\ref{fig:f1} shows corresponding results.  We find that the angular power spectrum estimated from the simulated MWA data is in good agreement with the model ( $\delta  \le \pm 20\%$, middle sub-panel) in the range $40 < \ell \leq 800$. The percentage deviation $\delta$ (in the middle sub-panel) increases slightly for  the lower and the higher $\ell$ values due the convolution and the system noise respectively. We also see that the deviation remains within the 1$\sigma$ region for the entire  $\ell$ range considered here. The deviation at low-$\ell$ is primarily due to the convolution with the kernel $\tilde{K}(\U_g,\U,\nu)$ (eq.~\ref{eq:c3}) and we expect the convolution to become important at  $\ell < 180^{\circ}/\theta_{\rm FWHM} \sim 36$. This is also seen in lower sub-panel, where the overlap of $|\tilde{w}_a^2(\ell)|^2$ between four low-$\ell$ bins are significant ($>0.5$). This  falls  below $|\tilde{w}_a^2(\ell)|^2 \sim 0.5$ at larger $\ell(>100$) bins. The overlap of  window functions increases significantly as compared with wide window functions (Figure~\ref{fig:fwide}) in the low-$\ell$ range. This is one of the downside of narrow window functions as it increases the correlation in the $\ell < 100$ bins significantly. Considering PC=ALL, we see that the behaviour of the estimated $\cl$ is quite similar to that for a single pointing.  Considering the errors,  we find that these are reduced to some extent at large $\ell$  ($\ge 200$) where the  system noise is expected to make a significant contribution to the total errors. 

Middle panel of Figure~\ref{fig:f1} is very similar to the left panel, except that we have now estimated $C_{\ell}$ for TC=C (hatched circle in Figure~\ref{fig:centers})instead of TC=H.  Considering TC=H,   this has $\delta_c=-26.7^{\circ}$ which matches  the declination $(\delta_p=-26.7^{\circ}$) of all the pointing in the drift scan observation considered here (Figure \ref{fig:centers}). In contrast, TC=C  has $\delta_c=-31.7^{\circ}$ which is $5^{\circ}$ below the drift scan track (Figure \ref{fig:centers}) and which does not coincide with any of the PC. We first consider a single pointing PC=16. Although the TC is shifted $5^{\circ}$ below  the PC, we find that estimated $C_{\ell}$ is in reasonably good agreement with the input model (upper sub-panel) in the range $50 < \ell < 600$. The percentage deviation $\delta$ is smaller than $\pm 20\%$ in this range except a single data point at $\ell \approx 300$. As mentioned earlier, this particular $\ell$-bin is rather sparsely sampled by the MWA baseline distribution (right panel of Figure~\ref{fig:bl}).   We find a noticeable improvement in the match when we  consider   PC=ALL,  particularly at large  $\ell$ range.  The estimated $\cl$  are now  in good agreement with $C_{\ell}^{M}$ in  the range $50 < \ell \le  800$. 

Overall the primary objective of using a narrow tapering function is to suppress foreground sources that are close to the main lobe of the primary beam. This significant decrease in $\thf$ increases the error in the estimated $\cl$ when compared with the wide tapering functions (Figure~\ref{fig:fwide}). Increase in cosmic variance and widening of the window function (lower sub-panels of Figure~\ref{fig:f1}) are the primary reason for the increase in error. This effects are prominent in the low-$\ell$ range, whereas, at large $\ell$  ($\ge 200$) we don't see significant deterioration of the estimated $\cl$.

Left panel of Figure~\ref{fig:f1_snr} show a comparison between the SNR achieved considering  PC=16 and PC=ALL,  we have used TC=H for both. Considering PC=16 with $2 \sigma_s $ noise, we find  a maximum  SNR of $2.7$ and the SNR $>2.5$ for a large number of $\ell$ bins in the range $100 < \ell < 800$. Increasing the noise level to  $5 \sigma_s$, we have a maximum SNR of $1.7$ {\it i.e.} a factor of $1.6$ smaller.  The  degradation   in SNR is not uniform across all $\ell$. This is more noticeable at large $\ell$ which are system noise dominated,  whereas the  SNR does not decrease much at small $\ell$ which are cosmic variance dominated. 
 The SNR  values  increase moderately  when we consider PC=ALL.  We now have  a maximum SNR  of  $4 $  and $2.7$ for  $2 \sigma_s$ and $5 \sigma_s$ noise respectively.  Comparing PC=16 with PC=ALL,  we see that for $\ell < 90$ the SNR values do not improve much  primarily because  the errors are cosmic variance dominated. There is a relatively larger increase in SNR at larger $\ell$ which are system noise dominated.  In all cases we also notice a sudden dip in the SNR around  $250 < \ell < 350$. This can be attributed to  the small number of baselines in this $\ell$ range  compared to the other $\ell$ bins (right panel of Figure~\ref{fig:bl}). Comparing middle panel of Figure~\ref{fig:f1_snr} with that of  left panel, we see that the two are very similar. One would expect the signal to be suppressed by the primary beam pattern as the TC moves away from the PC {\it i.e.} $\bm{\chi}_p$ increases (eq.~\ref{eq:f9}). However, our results indicate that this effect is not very severe for the set of PC and TC considered here. Not only are we able to faithfully  recover $C_{\ell}^M$, we also do not notice any substantial degradation in the SNR when the TC is moved $5^{\circ}$ below the track of the drift scan observations.

To assess the effect of the size of the narrow tapering functions, we have repeated the simulations using  $\thf = 6.67^{\circ}$ (\ie $\theta_{\rm FWHM} = 6.67^{\circ}$). In this case we expect the estimated $C_{\ell}$ to match $C_{\ell}^M$ for even smaller values of $\ell$ as compared to $\thf=5^{\circ}$.  Right panel of Figure~\ref{fig:f1} shows the estimated  $\cl$  for TC=H.  Considering  PC=16, we see that the estimated $C_{\ell}$ is in good agreement with $C_{\ell}^M$ in the range $40 < \ell \leq 800$. The $\ell$ range over which the percentage deviation $\delta$ remain within  $\approx \pm 20\%$ does not differ much when compared with $\thf = 5^{\circ}$  (middle sub-panel). However the $\delta $ values are reduced, and these are within  $10\%$  throughout the range $80 \lesssim \ell \lesssim 800$. Considering  PC=ALL, we see that the behaviour of the estimated $\cl$ is quite similar to that for a single pointing with a  small but  noticeable improvement in  $\delta$ at low-$\ell$. Comparing  the statistical fluctuations for PC=ALL with those for PC=16, we find that these are reduced to some extent for  the noise dominated large $\ell > 200$ range. This is very similar to what we have found earlier for $\thf=5^{\circ}$ in left panel of Figure~\ref{fig:f1}. Right panel of Figure~\ref{fig:f1_snr} shows the  SNR for $\thf = 6.67^{\circ}$  and TC=H. The solid angle subtended by the tapering window function  scales  as $\thf^2$ as $\thf$ is increased from $5^{\circ}$ to $6.67^{\circ}$. 
Considering a uniform baseline distribution, we expect the cosmic variance to decrease by a factor $(0.75)^2$ or the  SNR is expected to increase by a factor of $1.3$ provided it is entirely cosmic variance dominated. Comparing the SNR values we see that these do increase in comparison to  $\thf=5^{\circ}$ (left panel). However,  we find that the increase in the SNR is quite small ($\le 10 \%$)  in the $\ell$ range  $< 90$ which is expected to be cosmic variance dominated. In contrast,  we find  up to $75 \%$ increase in the SNR for  some of the large $\ell$ bins where  the system noise  and the cosmic variance  are both important.
We interpret this behaviour of the SNR as arising due to the  the non-uniform baseline distribution.This is also consistent with our finding from wide window functions. In addition to TC=H shown here, we have also considered TC=C for which the  results are not shown here. 
Overall, the results for $\thf=6.67^{\circ}$ are all qualitatively similar to those for  $\thf=5^{\circ}$, with the difference that the values of  $\delta$ decrease and SNR increase to some extent for  $\thf=6.67^{\circ}$.

 The fact that the SNR increases with $\thf$ may lead one to conclude that we should choose a large value for $\thf$. For example the MWA primary beam  has  $\theta_{\rm FWHM} \sim 23^{\circ}$, and we could possibly consider choosing  wider tapering function would have a FWHM comparable to $25^{\circ}$ (shown in Figure~\ref{fig:fwide}). However, it is also important to recollect that the main purpose of tapering the sky response is to cut down the foreground contamination from bright sources in the telescopes field of view, and   a small value of $\thf$ would provide better control over the foreground contamination. Unfortunately, this would reduce the accessible $\ell$ range and also the SNR values. To some extent, it  is possible to compensate for this loss in SNR by considering multiple TC within the telescope's field of view. The $C_{\ell}$ estimated for the different TC are expected to be independent provided that their angular separations are sufficiently large compared to  the sky response of the tapering window function.  It is then possible to increase the SNR by combining the $C_{\ell}$ estimated for the different TC.  
 In the next sub-section we have considered this  for $\thf=5^{\circ}$, the different TC here (Figure \ref{fig:centers}) are at an angular separation of $5^{\circ}$ which matches the FWHM of the tapering window function.

\begin{figure*}
    \centering
    \includegraphics[scale=0.33]{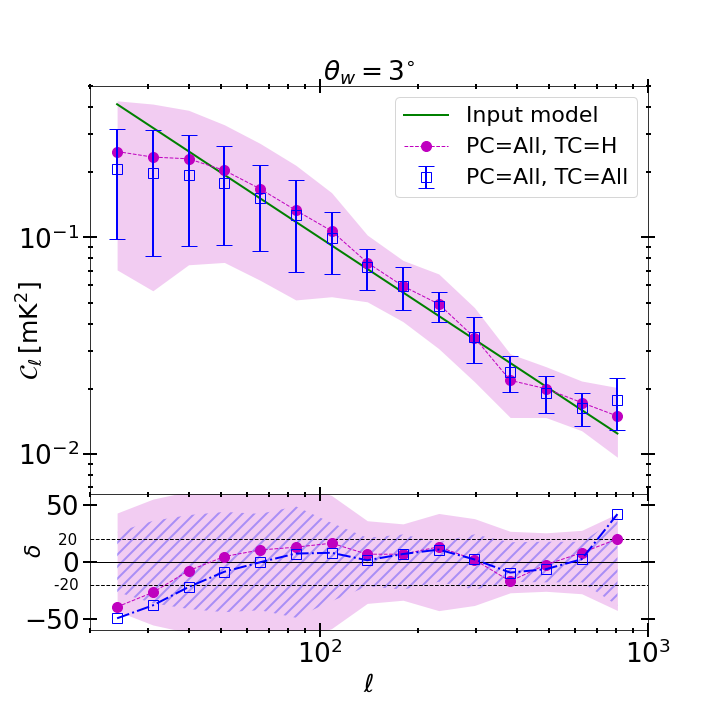}
    \includegraphics[scale=0.33]{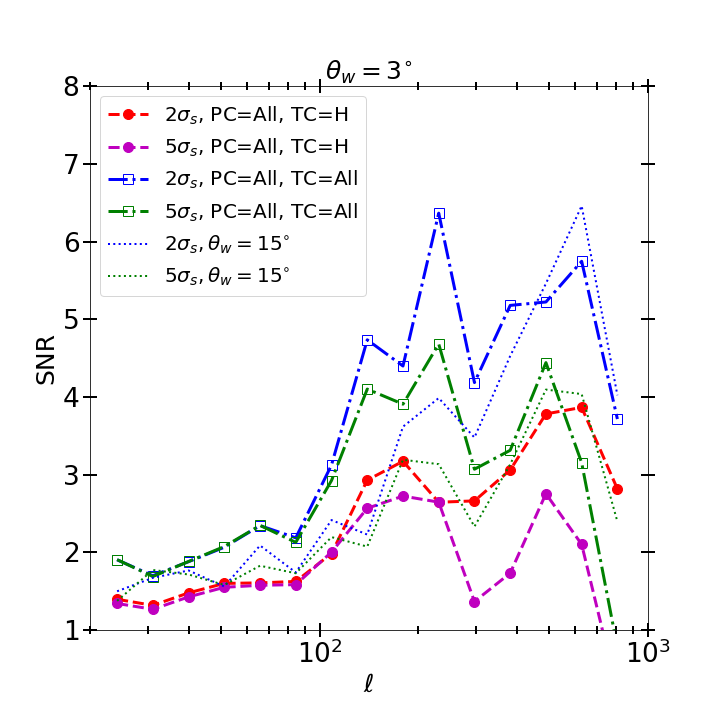}
    \caption{Left upper panel shows $\cl$ estimated for PC=ALL using TTGE with $\theta_w=3^{\circ}$.  The solid line shows the input model (eq.~\ref{eq:cl_1}) used for the simulations which have a noise level of $2 \sigma_s$.  50 realizations of the simulations were used to estimate the mean and $1 \sigma$ errors shown here.  The  magenta circles and the shaded region show the results for TC=H whereas the blue squares and error-bars correspond to TC=ALL \ie all the tracking centers combined. Considering the  $\cl$ estimates in the left upper panel,  the left lower panel shows the corresponding percentage deviation  relative to  the input model, the shaded and hatched regions show the $1\sigma$  statistical fluctuations for TC=H and TC=ALL respectively. The right panel show a comparison of the signal-to-noise-ratio (SNR) for the different cases considered here. In addition to the $2 \sigma_s$ noise level shown in the left panel, here we have also considered a noise level of $5 \sigma_s$. We also show the SNR achievable using TTGE with $\theta_w=15^{\circ}$ (in thin dotted lines, see Figure~\ref{fig:fwide}).}
    \label{fig:fc1}
\end{figure*}

\subsection{Combined tracking}\label{subsec:ct}
\label{sub:3}
We have estimated $C_{\ell}$ for all the 15 TC shown in  Figure \ref{fig:centers} and combined these  using  eq.~(\ref{eq:tc_avg}) to  obtain $C_{\ell}$ corresponding to PC=ALL, TC=ALL for which the results are shown in Figure~\ref{fig:fc1}. We have considered  $\thf=5^{\circ}$, and used $50$ realisations of the simulations to estimate the mean $C_{\ell}$ and $1 \sigma$ errors shown here. 
The upper left panel  shows $C_{\ell}$ for a system noise level of $2 \sigma_s$,  here $C_{\ell}$ for  PC=ALL, TC=H  (Figure \ref{fig:f1}) and $C^M_{\ell}$ are also shown for comparison. We see that the two estimates of  $C_{\ell}$  are in reasonable agreement across  most of the $\ell$ range, however the $C_{\ell}$ values  for TC=ALL are somewhat smaller  than those  for TC=H at $\ell \le 100$. The $1 \sigma$ error bars are smaller for TC=ALL in comparison to TC=H across the entire $\ell$ range.   The lower left panel of Figure~\ref{fig:fc1} show $\delta$ and the respective  $1 \sigma$ error-bars. For both TC=H and TC=ALL,  we see that the condition  $(\mid \delta  \mid )  \le  20\%$ is satisfied across the $\ell$ range $40 < \ell < 700$. In both cases the deviations from $C_{\ell}^M$ are negative for $ \ell \le 100$, and  the magnitude of the deviations are larger for TC=ALL. This is possibly because for TC=ALL we have (PC, TC) combinations like (33, A) and (1, E) (Figure \ref{fig:centers})  where the angular separations are quite large $(\mid \bm{\chi}_p \mid  \sim 15^{\circ})$, whereas the angular separations are restricted within $10^{\circ}$ for TC=H.

The right panel of Figure~\ref{fig:fc1} shows the SNR values for PC=ALL, TC=ALL considering two different  noise levels ($\sigma_n=2 \sigma_s, 5 \sigma_s$), the corresponding results for PC=ALL, TC=H  are also shown here for comparison. Considering the  single TC (=H) first, we see that for $\ell \le 100$ the SNR values do not change when $\sigma_n$ is varied,  indicating that the errors in this $\ell$ range are cosmic variance dominated. At  larger $\ell$,  
the errors are a combination of the cosmic variance and the system noise  and  the system noise contribution increases with $\ell$. When $\sigma_n$ is varied, there is a small, but noticeable,  change in the SNR for $100 < \ell <300$, whereas the  SNR values increase by a factor of $1.5-2$ in the $\ell$ range $\ell \ge 300$ which is system noise dominated. Considering TC=ALL next, we see that the same pattern is repeated \ie when  $\sigma_n$ is varied the
 SNR  does not change for  $\ell \le 100$, etc. However, the main difference is that the overall SNR values are larger for TC=ALL in comparison to TC=H. We see that this increase  depends  on the $\ell$ bin, and it has a value  $\sim 1.5$ on average.  For $\sigma_n=2 \sigma_s$, we find a maximum increase of $\sim 2.5$ around $\ell \approx 250$. Overall, we see that all the SNR curves are somewhat similar and they are all  related to the way in which  the number of baselines in each bin varies with $\ell$  (Figure \ref{fig:bl}). 
 
 Ideally we expect the SNR values to  increase  by a factor of $\sqrt{N_{\rm TC}} \approx 3.9$ when the measurements from $N_{\rm TC}=15$ TC are combined. This is under the assumption that 
  measurements of $\cl$ from different TC are  independent. However, we only find an increase of $1.5$ on average, and even the maximum increase is less than $3.9$. This deviation indicates that the  $\cl$ estimated at the different TC are correlated. The overlap between the tapering window functions for adjacent TC is one of the factors responsible for  this correlation. To test this we have considered $\thf=6.67^{\circ}$ for which there is a larger overlap between the window functions for adjacent TC. Combining the different TC, we find that the increase in the SNR for $\thf=6.67^{\circ}$   is smaller in comparison to that for $\thf=5^{\circ}$. This is consistent with the picture that a part of the correlation arises due to the overlap of the tapering window functions for adjacent TC. 
  In addition to this,  the fact that the $\cl$  for the different TC are all estimated from the same sparsely sampled visibility data also  contributes  to the correlation. 
  Recollect that in Section~\ref{sub:1} we have used one wide tapering function at TC=H and combined signals from 33 PC. In this section, measurements from fifteen small tracking centers (TC=ALL) with PC=ALL are combined to estimate $\cl$. The first approach is beneficial for a situation where we do not have any residual foregrounds left with in the main lobe of the telescope beam and all of the FoV can be used for signal estimation. Whereas, if unwanted contamination remains within the FoV of the telescope's beam, one may use restricted regions of the FoV to estimate signal. Right panel of Figure~\ref{fig:fc1} also show the SNR achievable using  $\thf=25^{\circ}$ with PC=ALL (in thin dotted lines). Considering  noise $\sigma_n=2 \sigma_s$, we find that the SNR for $\thf=25^{\circ}$ remains smaller than TC=ALL with $\thf=5^{\circ}$ except at the large $\ell$ values. Similar trend is also noticeable for  $\sigma_n=5 \sigma_s$. This primarily shows that both the approaches show similar performance at the scales much smaller than the tapering window where the system noise dominates. Whereas at the large scales \ie at the smaller $\ell (\leq 400$), TC=ALL performs noticeably better than wide tapering function. The cosmic variance at the small $\ell$ values can be reduced when we combine ALL tracking centres and we expect the SNR to increase in these $\ell$ ranges, as shown in the right panel of Figure~\ref{fig:fc1}. The TTGE is expected to perform better for an uniform dense baseline distribution and this trend in SNR values could arise due to the sparse distribution of MWA baselines. To demonstrate this, we have performed the same analysis with uniform baseline distribution in Appendix~\ref{sec:app2}. We found that with an uniform dense baseline distribution, both the approaches discussed in Section~\ref{sub:1} and ~\ref{sub:3} perform comparably (see Figure~\ref{fig:funi}).

\section{Summary and Discussion}\label{sec:summary}
Drift scan observations provide an economic and stable option with the broad sky coverage required for 21-cm intensity mapping experiments. In this paper we present an estimator, namely the Tracking Tapered Gridded Estimator (TTGE) and we aim  to quantify the power spectrum of the sky signal directly from the visibilities measured in radio interferometric observations in the drift scan mode. This estimator is tailored for telescopes which have a large fields-of-view  (\eg LOFAR, MWA, HERA, CHIME), however, for demonstration purposes, we adopted  MWA-like interferometric configuration in this paper.The TTGE allows us to track a fixed sky location (TC) as it drifts across the telescope's field of view. It  further allows us to taper the sky response to a small angular region around the TC so as to suppress the foreground contamination from bright sources located at large angular separations from the TC.

In this paper we have presented the mathematical framework for the TTGE, and validated this using simulations. The entire analysis here is restricted to a single frequency for which we have used the TTGE to estimate the angular power spectrum $C_{\ell}$. It is relatively straightforward to generalise this to multi-frequency simulations where we can estimate $C_{\ell}(\Delta \nu)$. However, substantially larger computational resources are needed for multi-frequency simulations, which is why we have restricted the analysis  here to a single frequency. Considering an input model $C_{\ell}^M \propto \ell^{-1}$,  we have simulated all-sky maps which were used to simulate drift scan observations with a MWA-like telescope.

The simulated drift scan track spans $16^{\circ}$ in RA with a fixed DEC of $-26.7^{\circ}$,  and  we have simulated the visibilities for  33 PC located at an equal interval of $0.5^{\circ}$ along the track  (Figure \ref{fig:centers}). In  Section~\ref{sub:1} we have considered wide tapering functions where the visibility data from a single PC is used to estimate $C_{\ell}$ for a single TC (=H)  whose sky position coincides with that of the PC. We find that TTGE is able to recover the input angular power spectrum reasonably well in the  $\ell$ range $30 < \ell \le 800$ \ie  the deviation  $\delta$  between the estimated $\cl$ and $C_{\ell}^M$ is  within $20 \%$  in this $\ell$ range (Figure \ref{fig:fwide}) . We next consider the possibility where the visibility data from the 33 PC are coherently combined and used to estimate $C_{\ell}$ for the same TC (=H). We expect the signal to be unchanged, however the system noise contribution is expected to be reduced.  We find that the estimated $C_{\ell}$ are not very different from those obtained with a single PC, and  TTGE is able to recover the input model reasonable well across the same $\ell$ range as for a single PC (Figure \ref{fig:fwide}). The SNR is however found to increases when we combine the data from 33 PC(Figure \ref{fig:fwide_snr}). This increase is only noticeable at large $\ell$ ($> 100$) where the statistical fluctuations are system noise dominated. At small $\ell$ which are cosmic variance dominated, the SNR does not change when we combine 33 PC.  This approach is beneficial for a situation where we do not have residual foregrounds with in the main lobe of the primary beam and all of the FoV can be used for signal estimation  (Appendix~\ref{sec:app1}). 
 
In Section~\ref{sub:2} we have considered the possibility that the foreground sources are located very close to or even within the main lobe of the primary beam and a narrow tapering function is required in order to mitigate the foregrounds. We have used  $\theta_w=3^{\circ}$ ($\theta_{\rm FWHM}=5^{\circ})$ for most of the following analysis. We have applied TTGE to this visibility data to estimate $C_{\ell}$ for 15 different TC  which are arranged in three rows at an  angular spacing  of $5^{\circ}$ (Figure \ref{fig:centers}).   We find that TTGE is able to recover the input angular power spectrum reasonably well in the  $\ell$ range $40 < \ell \le 800$ (Figure \ref{fig:f1}). The results are very similar to that of wide tapering functions except at the low-$\ell$ range where we find noticeable increase in percentage deviation $\delta$ due to cosmic variance and convolution effects of the window function.

It is possible to further reduce the statistical fluctuations by combining the $C_{\ell}$ values estimated at the different TC. In Section~\ref{sub:3} we assume that each TC provides an independent estimate of the signal and we expect the SNR to  increases  by a factor  of $\sqrt{N_{\rm TC}}$. Here we have combined the $C_{\ell}$ values estimated at the 15 TC (Figure \ref{fig:centers}) to obtain the combined $C_{\ell}$  (PC=ALL,TC=ALL),  and compared the results with those for a single TC (PC=ALL,TC=H). Considering (PC=ALL,TC=ALL) we find (Figure \ref{fig:fc1})  that the estimated $C_{\ell}$ is in reasonably good agreement with the input model over the $\ell$ range $40 < \ell < 700$ which is comparable to  that for a single pointing (PC=ALL,TC=H). Considering the SNR, we find that this increases by a factor $\sim 1.6$ on average  when we combine all the TC.  The actual increase in SNR varies across the $\ell$ bins, however it is less than $3.9$ which is expected if the $C_{\ell}$ estimates at the different TC were all independent. We attribute this difference to correlations in the $C_{\ell}$ values estimated from adjacent TC. Such correlations can arise from the overlap between the tapering window functions at adjacent TC, and also from the sparse distribution of the MWA baselines (Appendix~\ref{sec:app2}).

In conclusion, we have validated the TTGE as a power spectrum estimator for drift scan observations with wide filed radio interferometric arrays. It allows us to combine the visibility data from multiple PC to estimate the power spectrum of the  signal from a small sky region centered around a fixed TC.  Although each such power spectrum estimate (from narrow tapering functions) will only quantify  the signal from a small region of the sky, we can cover the entire angular footprint of the wide field drift scan observation using a population of TC and combine the results from the individual TC. In real observations, the locations  of these TC and the angular width of the tapering function would be decided based on multiple considerations. The TC would be placed so as  to avoid foreground contribution from bright sources.  A small  value of $\thf$ would reduce the foreground contamination, however this would be at the expense of reducing the available $\ell$ range and also the SNR (Figure \ref{fig:f1_snr}).  The value of $\thf$ would have to be optimised  so as to balance  these competing factors. The value of the weights  $s_p$ for the different PC is another quantity which needs to be decided. For the analysis presented here,  we have used equal weights  for all the 33 PC. In real observations  however, considering  any particular TC it will be necessary to suitably choose the weights so as to cut-off the contributions from the PC which are located at large angular separations from the TC.  We propose to address these issues and apply the TTGE to observational data in future work.

\section*{Acknowledgements}
S. Chatterjee is supported by the Department of Atomic Energy, Government of India, under project no. 12-R\&D-TFR-5.02-0700.
SB would like to acknowledge funding provided under the MATRICS grant SERB/F/9805/2019-2020 of the Science \& Engineering Research Board, a statutory body of Department of Science \& Technology (DST), Government of India.

\section*{Data availability}
The data underlying this paper will be shared on reasonable request
to the corresponding author


\bibliographystyle{mnras}
\bibliography{ref_all}
\vspace{1cm}
\appendix{\bf{APPENDIX}}

\section{A simple visibility correlation estimator} \label{sec:app1}
The multi-frequency angular power spectrum $C_{\ell}(\nu_a, \nu_b)$ defined in eq.(\ref{eq:cl}) jointly characterizes 
the  angular ($\ell$) and frequency dependence of the statistical properties of the sky signal. 
The observed visibilities $\V(\u_n,\nu)$ are related to $C_{\ell}(\nu_a, \nu_b)$ through the two visibility correlation 
which, can be written as
\begin{eqnarray}
&& \langle {\V}(\u_n, \nu_a)
\,{\V}^{*}(\u_m,\nu_b)\rangle= 
\dBdTnu^{2} 
 \int \, d^2 U^{'}  \, 
  \tilde{a}\left(\u_n-\u^{'},\nu_a\right)  \times \, \nonumber  \\
 && \quad  \tilde{a}^{*}\left(\u_m-\u^{'},\nu_b\right)\, 
C_{2 \pi U^{'}}(\nu_a, \nu_b) \,.
\label{eq:ac3}
\end{eqnarray} 
Here the two visibilities are strongly correlated only if they correspond to the same baseline and the correlation decreases as the difference increases. For this demonstration we restrict our analysis to the situation where the two baselines are the
same.  Further, in the subsequent discussion we also ignore the frequency dependence.  The two
visibility correlation can then be expressed as 
\begin{equation}
{V}_2(\u_n)  \equiv \langle {\V}(\u_n)
{\V}^{*}(\u_n)\rangle
\label{eq:ac4}
\end{equation}
and we have 
\begin{eqnarray} 
 {V}_2(\u_n) = \dBdTnu^{2}  \int d^2 {\u}^{\prime} \,|\tilde{a}\left(\u_n - 
 {\u}^{\prime} \right)|^2\, C_{2\,\pi\, U^{\prime}}. 
\label{eq:ac5}    
\end{eqnarray}
where we do not explicitly show $\nu$ as an argument in any of the
terms. 

\begin{figure}
    \hspace{-0.50cm}
    \includegraphics[scale=0.40]{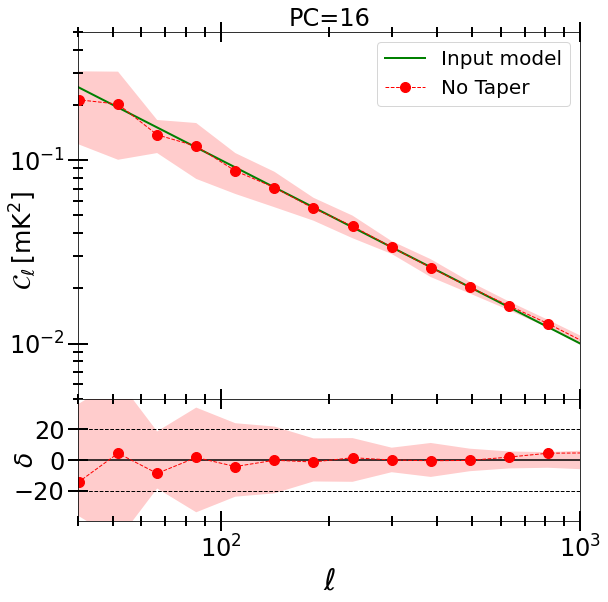}
    \caption{The upper panel shows $\cl$ estimated using a simple visibility correlation estimator.  The solid line shows the input model (eq.~\ref{eq:cl_1}) used for the simulations which 
    have a no added noise.  50 realizations of the simulations were used to estimate the mean and $1 \sigma$ errors shown here.  The  red circles and the shaded region show the results for PC=16. Considering the  $\cl$ estimates in the upper panel,  the  lower panel shows the corresponding percentage deviation  relative to  the input model, the shaded region shows the $1\sigma$  statistical fluctuations.}
    \label{fig:v2}
\end{figure}

At large baselines it is possible to approximate the convolution
in (eq. \ref{eq:ac5}) as 
\begin{eqnarray} 
 {V}_2(\u_n) =
\dBdTnu^{2} 
\left[ \int d^2 {\u}^{\prime} \,|\tilde{A}\left(\u_n -  
 {\u}^{\prime} \right)|^2\, \right] C_{\ell}. 
\label{eq:ac6}    
\end{eqnarray}
with $\ell = 2 \pi U_n$. We use eq.~\ref{eq:ac6} to estimate the $\cl$ from our simulation. For further details see \citet{Ali2014}. This simple visibility correlation estimator (VCE) does not taper the sky response as we have discussed earlier in case of TTGE. 
Here we consider a single frequency channel to validate our formalism for the VCE,  however it will be straightforward to generalize for multi-frequency observations \citep{Ali2014}. We use 50 realizations of the model angular power spectrum eq.~(\ref{eq:cl_1}) to generate the visibilities for validation. We consider a situation where we use the visibility measurements from a single pointing center (PC=16). The upper panel of Figure~\ref{fig:v2} shows a comparison between the input $\cl^M$ (in green solid-line) and estimated $\cl$. The red filled circles show the mean and the shaded region shows the 1$\sigma$ rms fluctuations of the $\cl$ estimated from 50 realizations of simulated visibilities  using the VCE.  It is worth mentioning there is no noise associated with this simulations. The lower panel of Figure~\ref{fig:v2} shows the  percentage deviation $\delta  =100 \, \% \times \, (\cl -\cl^M)/ \cl^M)$  and the red shaded  region shows the 1$\sigma$ statistical fluctuations for the same. The dashed horizontal lines mark the values $\delta   =   \pm 20 \%$. We find that the angular power spectrum estimated from the simulated MWA data is in good agreement with the model ( $\delta  \le \pm 20\%$ ) in the range $\ell \leq 800$. 
 The deviation at small $\ell$ is due to the approximation used in eq.~\ref{eq:ac6} which becomes important at small baselines.

It is very difficult to model and subtract out bright point sources located at a considerable angular distance from the phase center. This is due to ionospheric fluctuations and also the inaccuracy of the primary beam pattern at the outer region of the main lobe. Several studies (\eg \citealt{Thyagarajan2015,Pober2016}) have shown that such sources  contaminate the  cylindrical 21-cm power spectrum $P(k_{\perp}, \, k_{\parallel})$ at the higher $k_{\parallel}$ modes. With the following exercises, we aim to demonstrate that the  TTGE suppresses the contribution from sources far away from the phase center by tapering the sky response with a tapering function which falls off faster than the  antenna primary beam (PB). To demonstrate that we simulate the extragalactic  Point Sources (EPS) that are expected to dominate the 150 MHz sky.  We simulate 30 realizations of all-sky EPS distribution using the method described in \citet{Chatterjee2021} (see Section 2.2). We model the differential source count of the sources using the fitting formula given by \citet{Franzen2019}. 

\begin{figure*}
    \hspace{-0.5cm}
    \includegraphics[trim={3.5cm 0 3.5cm 0},clip,scale=0.45]{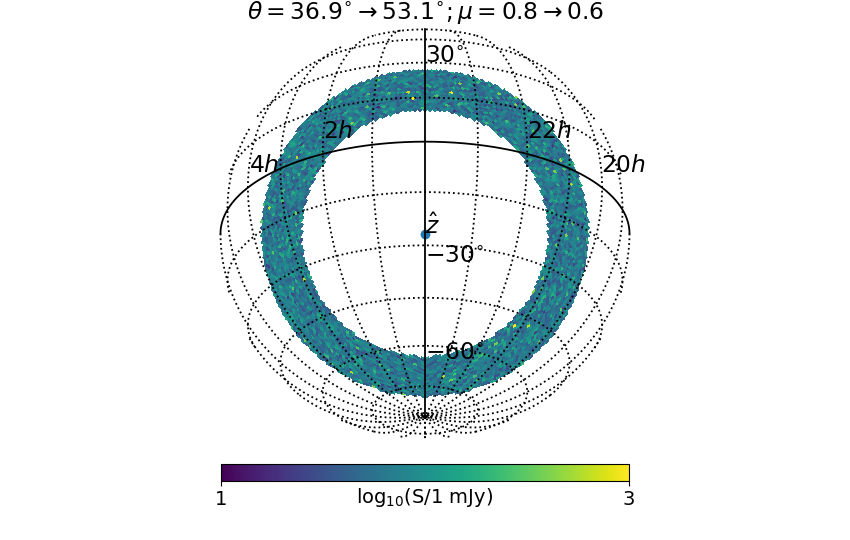}
    \hspace{-0.99cm}
    \includegraphics[trim={3.5cm 0 3.5cm 0},clip,scale=0.45]{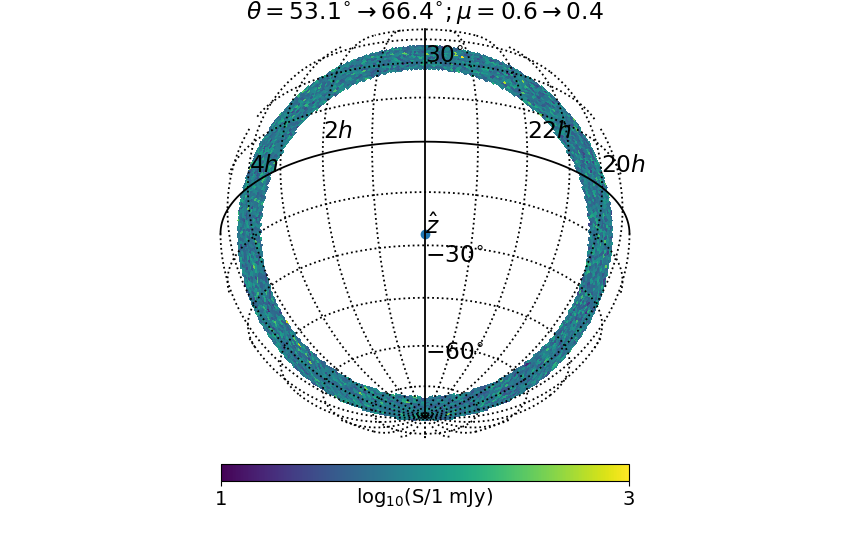}
    \hspace{-0.99cm}
    \includegraphics[trim={3.5cm 0 3.5cm 0},clip,scale=0.45]{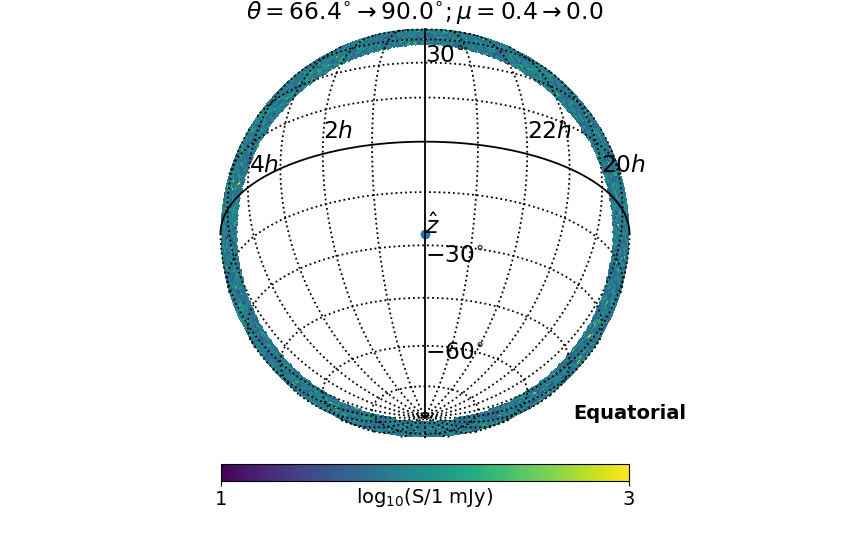}

    \hspace{-0.9cm}
    \includegraphics[trim={0 0 0 0},clip,scale=0.4]{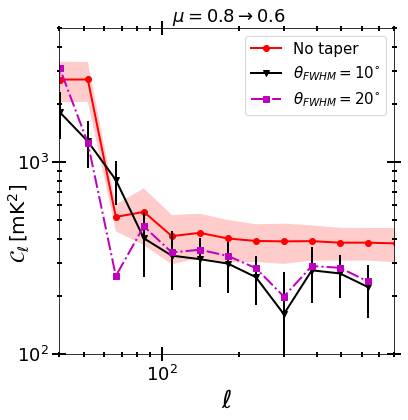}
    \hspace{-0.1cm}
    \includegraphics[trim={0 0 0 0},clip,scale=0.4]{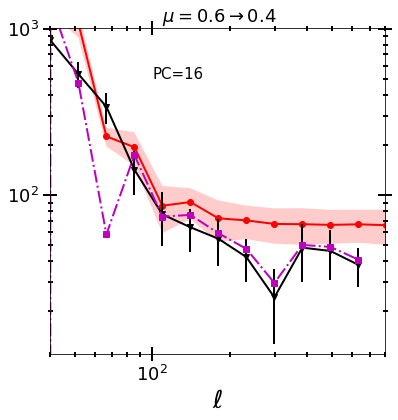}
    \hspace{-0.1cm}
    \includegraphics[trim={0 0 0 0},clip,scale=0.4]{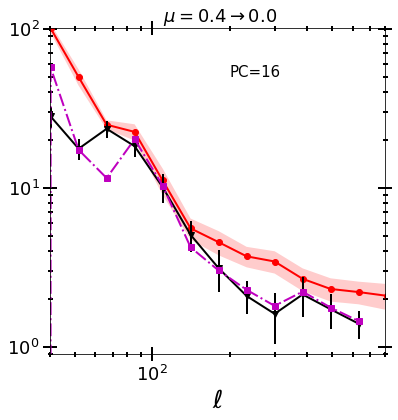}
    \caption{Upper panels of the figure show simulated point source distributions restricted to three different annular regions. Angular dimensions of theses annular regions are mentioned in the figure. $\hat{\mathbfit{z}}$ here denotes the pointing direction of the telescope. Bottom panels show a comparison between the $\cl$ estimated using a VCE (in red) and TTGE for two different tapering windows with $\thw = 12^{\circ} \, {\rm and} \, 6^{\circ}$ \ie $\thf = 20^{\circ} \, {\rm and} \, 10^{\circ}$ (in black and magenta).}
    \label{fig:af2}
\end{figure*}

As it is particularly difficult to remove point source contributions from the side lobes and that are closer to the horizon, here we have considered three different annulus regions in the upper-half of the sky, that are beyond FWHM of the PB. The three regions are chosen to be non-overlapping between $\mu (\equiv\cos \theta) = 0.8 -0.6, 0.6-0.4 \, {\rm and} \, 0.4-0$ (shown in the upper panels of Figure~\ref{fig:af2}), where $\theta$ is the zenith angle. We have considered the visibility contribution coming from these annulus regions only and masked rest of the sky. The results are shown in Figure~\ref{fig:af2}. For this whole demonstration we have considered PC=16 and TC=H, which coincides with the PC in consideration (see Figure~\ref{fig:centers}).

The upper panels of Figure~\ref{fig:af2} shows one realization of the EPS distribution within the three annulus region considered here (extent of these regions are mentioned in figure). The three lower panels show the $\cl$ estimates. Here, we have ignored the system noise contributions to the visibilities. Considering the lower-left panel, red filled circles show the $\cl$ estimates using the VCE (Eq.~\ref{eq:ac6}) whereas the shaded region shows the $1\sigma$ error on the in the $\cl$ estimates. The filled triangles and squares show the $\cl$ estimates using TTGE. With VCE, in the $\ell > 90$ range, the EPS contribution to the $\cl$ estimates remains flat and does not change with increasing $\ell$ (in red circles). $\cl$ estimates with TTGE also show similar behaviour. However the amplitudes of the $\cl$ estimates are reduced for TTGE. Considering $\thw = 12^{\circ}$ (\ie $\thf = 20^{\circ}$) we see that the EPS contribution (shown by magenta squares) in $\ell > 90$ range have been reduced by $\sim 40 \%$ when compared with VCE. For a smaller tapering window with $\thw = 6^{\circ}$ (\ie $\thf = 10^{\circ}$) the EPS contribution (in black triangles) is further suppressed. In the middle and right panels we consider the EPS contribution further away from the main lobe of the PB. In the middle panel we see similar suppression (\ie $\sim 40 \% $) in the $\cl$ estimates with TTGE when compared with VCE. The lower-right panel shows the EPS contribution coming from the region close to horizon. We also find noticeable suppression in the EPS contribution to the $\cl$ that is close to $\sim 30 \%$ with TTGE. It is worth mentioning that the relative suppression between tapering widows with $\thf = 20^{\circ} \, {\rm and} \, 10^{\circ}$ successively reduce, as we move EPS further away from the main lone. This exercise demonstrate the effectiveness of the TTGE in suppressing the foreground contribution coming from the side lobes, even regions closer to the horizon.  

\section{Effect of baseline coverage} \label{sec:app2}
\begin{figure}
    \hspace*{-0.95cm} 
    \includegraphics[scale=0.35]{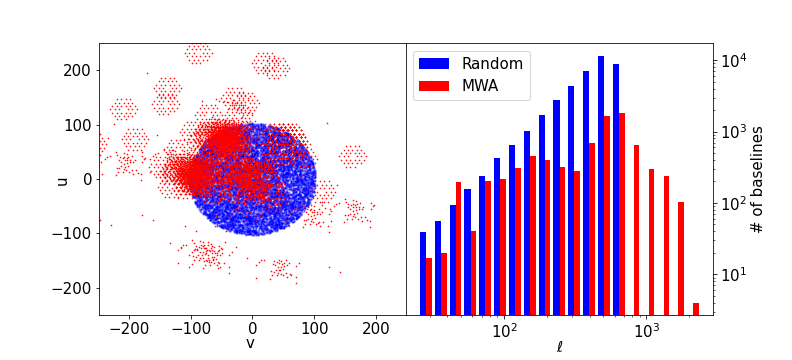}
    \caption{The left panel shows the MWA and uniform baseline $(uv)$ distribution in wavelength units.  Dividing the baselines into logarithmic bins of equal spacing, the  right panel shows the number of baselines in each  $\ell$ bin. }
    \label{fig:bluni}
\end{figure}

In this section, we consider the possibility of using TTGE for a radio interferometric array with uniform baseline($uv$) distribution to estimate the angular power spectrum.  Here, we have generated 10 independent realizations of the sky signal. The simulations were carried out in exactly the same way as described in Section ~\ref{sec:simulation} using the uniform $uv$ distribution (Figure~\ref{fig:bluni}). Visibilities were generated for 33 pointing centers and we have increased the baseline number to five times of the number of MWA baselines. However we have restricted the baseline in the range $\vert u, v \vert \leq 100$. This is primarily to show the effectiveness of TTGE for an uniform and dense $uv$ distribution. We expect this scenario to perform significantly better than the MWA-like non-uniform and sparse $uv$ distribution. Figure~\ref{fig:bluni} shows a comparison of $uv$ distribution for these two scenarios. 

\begin{figure*}
    \hspace{-1cm}
    \includegraphics[scale=0.33]{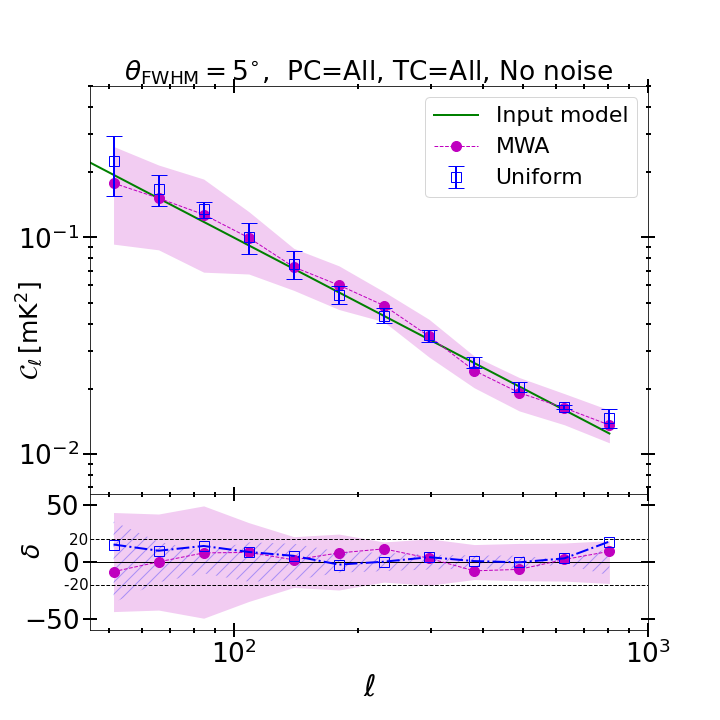}
    \hspace{01cm}
    \includegraphics[scale=0.33]{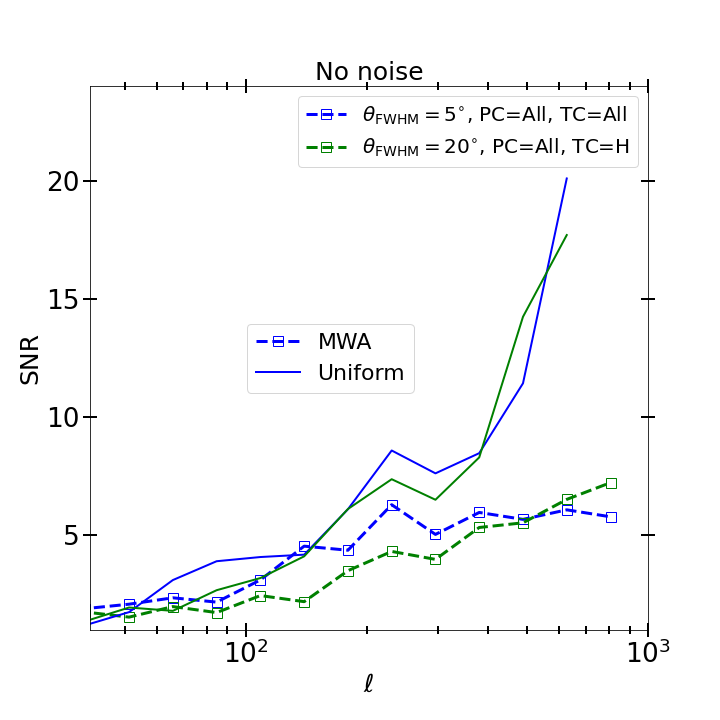}
    \caption{Left upper panel shows $\cl$ estimated for PC=ALL and TC=ALL using TTGE with $\thf=5^{\circ}$.  The solid line shows the input model (eq.~\ref{eq:cl_1}) used for the simulations which do not have noise.  The  magenta circles and the shaded region show the results for MWA whereas the blue squares and error-bars correspond to uniform baseline distribution. Considering the  $\cl$ estimates in the left upper panel,  the left lower panel shows the corresponding percentage deviation  relative to  the input model, the shaded and hatched regions show the $1\sigma$  statistical fluctuations for MWA and uniform baseline distribution respectively. The right panel show a comparison of the SNR for the different cases considered here (in dashed lines). We also show the SNR achievable using with uniform dense baseline distribution (in solid lines). }
    \label{fig:funi}
\end{figure*}

Left panel of the Figure ~\ref{fig:funi} shows a comparison between the estimated $\cl$  using MWA and uniform $uv$ distribution. For this comparison we have set $\thf=5^{\circ}$, PC=ALL, TC=ALL and  ignored the system noise contribution. We see that the two estimates of  $\cl$ the $1 \sigma$ error bars are significantly smaller for uniform distribution (hatched region in lower panel) in comparison to MWA (shaded region in lower panel) primarily because of large number of baselines used in case of uniform baselines. Also, the fractional deviations for uniform $uv$ distribution are small as compared with the MWA configurations. In right panel of Figure~\ref{fig:funi}, we compare the SNR values for two different cases (a) $\thf=5^{\circ}$ with TC=ALL and (b) $\thf=20^{\circ}$ with TC=H. The dashed lines are for MWA, and the corresponding SNR for uniform $uv$ distribution are shown in solid lines. Our aim is to study the relative performance of TTGE for a wide tapering function (as discussed in Section~\ref{sub:1}) and combined tracking with multiple small tapering functions (as discussed in Section~\ref{sub:3}) considering different baseline distributions (\ie uniform and MWA). Considering uniform $uv$,  we find that the SNR values for $\thf=5^{\circ}$ with TC=ALL and $20^{\circ}$ with TC=H, are comparable throughout the $\ell$ range considered here.  However for MWA baselines, we find that in $\ell$ range $40 < \ell < 500$, $\thf=5^{\circ}$ with TC=ALL have larger SNR compared to $\thf=20^{\circ}$ with TC=H. The relative performance is reversed in the $\ell > 500$. These results are different from what we found in case with uniform $uv$ distribution. This shows that the baseline distribution plays a crucial role in power spectrum estimation with TTGE. Comparing the SNR values, we conclude that the TTGE performs better for a uniform and dense $uv$ coverage.

\begin{figure*}
    \includegraphics[trim={0 0 0 0},clip,scale=0.4]{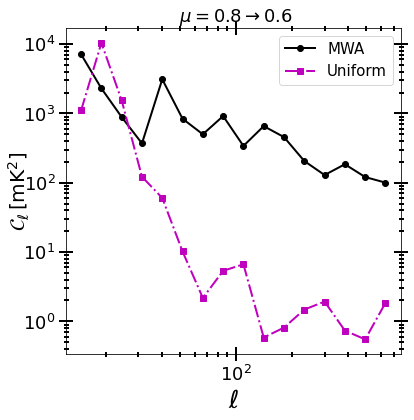}
    \hspace{-0.1cm}
    \includegraphics[trim={0 0 0 0},clip,scale=0.4]{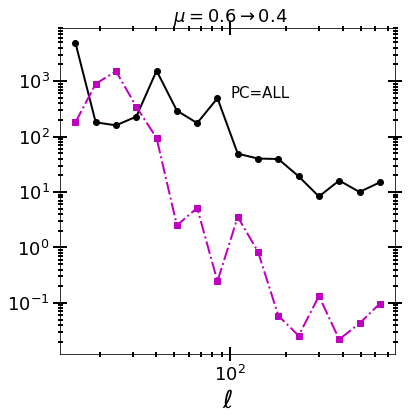}
    \hspace{-0.1cm}
    \includegraphics[trim={0 0 0 0},clip,scale=0.4]{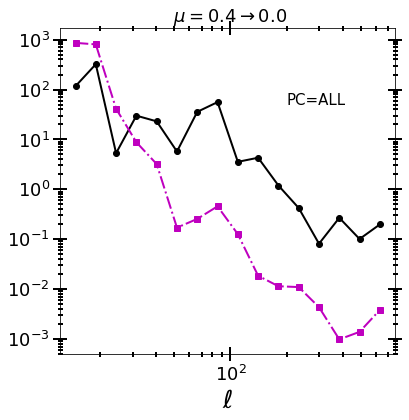}
    \caption{Three panels of the figure show the $\cl$ estimated using tapering window with $\thf = 20^{\circ}$, for simulated point source distributions restricted to three different annular regions. Angular dimensions of theses annular regions are mentioned in the figure. Circles (solid line) and squares ( dot-dashed line) show the results for MWA and uniform baseline distributions respectively.}
    \label{fig:af3}
\end{figure*}

Finally we show the results for the effect on the point source contributions from the side lobes and that are closer to the horizon in Figure~\ref{fig:af3}. Here we have considered same three annulus regions in the upper-half of the sky, that are beyond FWHM of the PB (shown in the upper panels of Figure~\ref{fig:af2}). We have considered the visibility contribution coming from these annulus regions only and masked rest of the sky. For this whole demonstration we have considered PC=ALL and TC=H with $\thf=20^{\circ}$.  Circles (solid line) and squares ( dot-dashed line) show the results for MWA and uniform $uv$ distributions respectively. The results are shown for a single realization of EPS simulation. In all the three panels we see that EPS contributions to the $\cl$ is significantly suppressed for uniform $uv$ distribution when compared with the results for MWA. This demonstrates the fact that considering TTGE an uniform and dense $uv$ coverage can suppress EPS contributions better than a sparse baseline coverage such as MWA. However as we have seen earlier even with a sparse baseline coverage like MWA, TTGE is able to suppress foreground contributions coming from side-lobes better than that of a simple VCE. 

\label{lastpage}

\end{document}